\documentclass[prd,amsfonts,noshowpacs,nofootinbib]{revtex4}
\usepackage{amsmath,graphicx}

\def\k{{\bf k}}

\begin{document}
\title{Trispectrum estimator
in equilateral type non-Gaussian models}
\author{Shuntaro Mizuno\footnote{shuntaro.mizuno@port.ac.uk}}
\author{Kazuya Koyama\footnote{Kazuya.Koyama@port.ac.uk}}
\affiliation{Institute of Cosmology and Gravitation, University of Portsmouth, Portsmouth PO1 3FX, UK.
}

\date{\today}
\begin{abstract}
We investigate an estimator to measure the primordial trispectrum
in equilateral type non-Gaussian models
such as k-inflation, single field DBI inflation and
multi-field DBI inflation models from Cosmic Microwave Background (CMB)
anisotropies. The shape of the trispectrum whose amplitude is not
constrained by the bispectrum in the context of effective theory of inflation
and k-inflation is known to admit a separable form of
the estimator for CMB anisotropies.
We show that this shape is $87 \%$ correlated with
the full quantum trispectrum in single field DBI inflation, while it is
$33 \%$ correlated with the one in
multi-field DBI inflation when curvature perturbation
is originated from purely entropic contribution.
This suggests that $g_{\rm NL} ^{equil}$, the amplitude of this particular
shape, provides a reasonable measure of
the non-Gaussianity from the trispectrum in equilateral non-Gaussian models.
We relate model parameters such as the sound speed, $c_s$ and the transfer coefficient
from entropy perturbations to the curvature perturbation, $T_{\mathcal{R} S}$
with $g_{\rm NL} ^{equil}$, which enables us to constrain model parameters in
these models once $g_{\rm NL}^{equil}$ is measured in WMAP and Planck.
\end{abstract}

\maketitle

\section{Introduction}
The statistical properties of primordial fluctuations provide crucial
information on the physics of the very early universe
\cite{Komatsu:2010fb, Smith:2009jr,
Senatore:2009gt, Fergusson:2010dm} (See \cite{Bartolo:2004if} for a review).
In the simplest single field inflation models where the scalar field has a canonical
kinetic term and quantum fluctuations are generated from the standard
Bunch-Davis vacuum, non-Gaussianity of the fluctuations is too small to
be observed even with future experiments
\cite{Acquaviva:2002ud, Maldacena:2002vr, Seery:2005wm}.
Thus the detection of
non-negligible departures from Gaussinaity of primordial fluctuations
will have a huge impact on the models of early universe.
So far, most of
the studies have focused on the leading order non-Gaussianity measured
by the three-point function of Cosmic Microwave Background
(CMB) anisotropies,
i.e. the bispectrum
\cite{Verde:1999ij,Wang:1999vf,Komatsu:2001rj}.
Especially, the optical method of extracting
the bispectrum from the CMB data has been sufficiently
developed
\cite{Komatsu:2003iq, Babich:2004gb, Babich:2005en, Creminelli:2005hu,
Creminelli:2006gc, Yadav:2007rk, Yadav:2007ny}
(for a more general approach, see
\cite{Fergusson:2006pr, Fergusson:2008ra,Fergusson:2009nv}).
However, future experiments
like Planck \cite{Planck} can also prove the higher order statistics
such as the trispectrum \cite{Hu:2001fa,Okamoto:2002ik,Kogo:2006kh}. The
trispectrum gives information that cannot be obtained from the
bispectrum \cite{Seery:2006js, Seery:2006vu, Byrnes:2006vq}.
In addition, it is possible that the trispectrum can be the
leading order non-Gaussianity,
that is, even if we do not detect the
bispectrum, this does not mean that the primordial fluctuations are
confirmed to be Gaussian.

In the local type non-Gaussian models \cite{Byrnes:2010em, Langlois:2010vx, Tanaka:2010km, Wands:2010af}, the primordial curvature
perturbation $\zeta$ is modeled as
\begin{equation}
\zeta =\zeta_{g} + \frac{3}{5} f^{local}_{\rm NL}
(\zeta_{g}^2 - \langle \zeta_{g}^2 \rangle) + \frac{9}{25} g^{local}_{\rm NL} \zeta_{g}^3
\end{equation}
where $\zeta_{g}$ is a Gaussian variable  \cite{Sasaki:2006kq}.
In this model, the bispectrum has maximum amplitude for the squeezed
configurations in the Fourie space where one of wavenumbers is small
compared with others. In these models, the trispectrum indeed gives very
interesting tests on multi-field inflation models where there are
several Gaussian variables $\zeta_g$ using the consistency relation
between the amplitudes of the bispectrum and trispectrum
\cite{Suyama:2007bg}. Also the trispectrum can constrain the cubic-order
non-linearities in primordial curvature perturbation, $g^{local}_{\rm
NL}$ that cannot be constrained by the bispectrum
measurements.
Estimators to measure the trispectrum in the local-type
non-Gaussianity have been developed and
the kurtosis based estimator \cite{Munshi:2009wy}
have been used to obtain constraints
on the amplitude of the trispectrum,
$-7.4 < g_{\rm NL} ^{local} / 10^5 < 8.2$ at $95 \%$
confidence level from the WMAP 5-year data \cite{Smidt:2010sv}.

There are another class of non-Gaussian models. A typical example is Dirac-Born-Infled (DBI) inflation
\cite{Silverstein:2003hf} whose non-Gaussian property
was extensively studied by
\cite{Alishahiha:2004eh, Chen:2004gc,Chen:2005ad,Chen:2005fe, Shandera:2006ax, Chen:2006nt,  Langlois:2008wt, Langlois:2008qf, Arroja:2008yy, Khoury:2008wj, Langlois:2009ej, Cai:2009hw}
(see also \cite{Koyama:2010xj, Chen:2010xk} for reviews).
In this model, like k-inflation
\cite{ArmendarizPicon:1999rj,Garriga:1999vw},
the inflaton field has non-canonical kinetic term and non-linear derivative interactions can give rise to large non-Gaussianity of quantum fluctuations.
For current observational constraints on DBI inflation see
\cite{Kecskemeti:2006cg,Lidsey:2006ia,Baumann:2006cd, Bean:2007hc, Lidsey:2007gq, Peiris:2007gz, Kobayashi:2007hm, Easson:2007dh, Lorenz:2007ze, Bean:2007eh, Bird:2009pq, Bessada:2009pe, Copeland:2010jt}.
In these models, the amplitude of the bispectrum has a peak typically for the equilateral configuration in the Fourie space. The shape of the trispectrum is more complicated. For example, for the bispectrum, the equilateral condition $k_1=k_2=k_3$ completely specifies the shape of the bisepctrum, but this is not the case for the trispectrum. The shape of the trispectrum has been analyzed in several inflationary models such as single field DBI inflation
\cite{Huang:2006eh, Arroja:2008ga, Arroja:2009pd, Chen:2009bc}, multi-field DBI inflation
\cite{Gao:2009gd, Mizuno:2009cv, Mizuno:2009mv, RenauxPetel:2009sj,
Gao:2009at} and the models motivated by effective theory
of inflation \cite{Senatore:2010jy, Bartolo:2010di}.

Regardless of these efforts, since the form of the
trispectrum is generally very complicated, estimators for the trispectrum in this class of
non-Gaussian models have not been implemented yet so far.
It was suggested that the form of trispectrum given by
\begin{equation}
T_{\zeta}(k_1,k_2,k_3,k_4) =\frac{g_{\rm NL}^{equil}}{k_1 k_2 k_3  k_4
 (\frac{k_1+k_2+k_3+k_4}{4})^5} {\cal P}_\zeta (k)^3\,,
\label{fiducial}
\end{equation}
represents the shape of the trispectrum in equilateral non-Gaussian models very well.
Here the trispectrum of the curvature perturbation is defined as
\begin{equation}
\langle \zeta(\k_1) \zeta(\k_2) \zeta(\k_3) \zeta(\k_4) \rangle
=(2 \pi)^3 \delta^3(\k_1+\k_2+\k_3+\k_4) T_{\zeta}(k_1,k_2,k_3,k_4),
\end{equation}
where ${\cal P}_\zeta (k)$ is the power spectrum given by
$\langle \zeta ({\bf k_1}) \zeta ({\bf k_2}) \rangle
= (2 \pi)^3 \delta^{(3)} ({\bf k_1} + {\bf k_2}) k_1^{-3} {\cal P}_\zeta (k_1)$. This trispectrum (\ref{fiducial}) appears in DBI inflation as a contribution from the
fourth order interacting Hamiltonian
(the ``contact interaction") \cite{Arroja:2009pd,Chen:2009bc}. In the effective theory of inflation, it was shown that the trispectrum of this shape can have the amplitude that is not constrained by the
bispectrum measurements \cite{Senatore:2010jy}.
In Ref.~\cite{Chen:2009bc} , it was suggested that this trispectrum can be used to
construct an estimator because by introducing the integral $1/M^n =(1/\Gamma(n))
\int^{\infty}_{0} t^{(n-1)} e^{-M t}$, this function is factorisable (see Appendix A).
Therefore, in this paper, we compare the shapes of trispectra in
single field and multi-field DBI inflation with Eq.~(\ref{fiducial})
based on a shape correlator introduced by Regan et.al \cite{Regan:2010cn} and investigate
whether the estimator constructed from the trispectrum (\ref{fiducial}) represents the
shapes of trispetrum in these models or not.

This paper is organized as follows. In section II, we review the shape correlator introduced by
Regan et al. \cite{Regan:2010cn}. In section III, we study the overlap
between the shape given by Eq.~(\ref{fiducial}) and trispectra in single
field and multi-field DBI inflation models.
In section IV, we give theoretical predictions for $g_{\rm NL} ^{equil}$ in some concrete
theoretical models. We conclude in section V. In Appendix A, we present the optimal estimator
using Eq.~(\ref{fiducial}) explicitly. In Appendix B, we summarise the shape function
of the reduced trispectra appeared in general single field
k-inflation models and give the shape
correlations among the representative shapes. In Appendix C,
we check the validity of our method to relate the amplitude of the estimator to the
theoretical predictions using the bispectrum.

\section{The shape correlator}

In this section, 
we review the shape correlator introduced by
Regan et al. \cite{Regan:2010cn}.

\subsection{Shape functions}
First we exploit the symmetry of the trispectrum to define the {\it reduced}
trispectrum as follows \cite{Hu:2001fa}. 
We rewrite the definition of the trispectrum as
\begin{eqnarray}
\langle \zeta({\bf k_1})  \zeta({\bf k_2})
 \zeta({\bf k_3})  \zeta({\bf k_4})\rangle_c
&=&(2 \pi)^3 \int d^3 K \bigl[
\delta({\bf k_1} + {\bf k_2} - {\bf K})
\delta({\bf k_3} + {\bf k_4} + {\bf K})
\bigl(\mathcal{T}_\zeta ({\bf k_1}, {\bf k_2}, {\bf k_3}, {\bf k_4};
{\bf K})\nonumber\\
&&+  \mathcal{T}_\zeta ({\bf k_2}, {\bf k_1}, {\bf k_3}, {\bf k_4};
{\bf K})+\mathcal{T}_\zeta ({\bf k_1}, {\bf k_2}, {\bf k_4}, {\bf k_3};
{\bf K})
+ \mathcal{T}_\zeta
({\bf k_2}, {\bf k_1}, {\bf k_4}, {\bf k_3};
{\bf K})\bigr)\nonumber\\
&& + ({\bf k_2} \leftrightarrow {\bf k_3}) +
({\bf k_2} \leftrightarrow {\bf k_4})
\bigr]\,.
\end{eqnarray}
Then we need to consider only the reduced trispectrum
$\mathcal{T}_\zeta$ from one particular arrangement of the vectors,
such as $\mathcal{T}_\Phi ({\bf k_1}, {\bf k_2}, {\bf k_3}, {\bf k_4};
{\bf k_{12}})$ with ${\bf k_{12}}={\bf k_1} + {\bf k_2}$
and form the other contributions by considering permutations. Here, for the later convenience,
we use the symmetrised reduced trispectrum
\begin{eqnarray}
\mathcal{T}_\zeta ^{sym} ({\bf k_1}, {\bf k_2}, {\bf k_3}, {\bf k_4};{\bf k_{12}}) &\equiv& \frac{1}{4}
\bigl[\mathcal{T}_\zeta ({\bf k_1}, {\bf k_2}, {\bf k_3}, {\bf k_4};{\bf k_{12}})+\mathcal{T}_\zeta ({\bf k_2}, {\bf k_1}, {\bf k_3}, {\bf k_4};{\bf k_{12}})
\nonumber\\
&&+ \mathcal{T}_\zeta ({\bf k_1}, {\bf k_2}, {\bf k_4}, {\bf k_3};{\bf k_{12}})+\mathcal{T}_\zeta ({\bf k_2}, {\bf k_1}, {\bf k_4}, {\bf k_3};{\bf k_{12}})\bigr]\,,
\label{sym_reducedtrispectrum}
\end{eqnarray}
and from now on we omit the superscript $sym$ for simplicity.

The reduced trispectrum
is a function of six variables.
We can choose them to be $(k_1,k_2,k_3,k_4,k_{12},\theta_4)$
where $\theta_4$ represents the deviation of the quadrilateral from planarity which is specified
by the triangle $(k_1, k_2, k_{12})$.
We find that in terms of these variables
$k_{14}=|{\bf k_1}+{\bf k_4}|$ is expressed as
\begin{eqnarray}
k_{14}^2 &=& k_1^2 + k_4^2 - \frac{1}{2 k_{12}^2}
(k_1^2 + k_{12}^2 - k_2^2)(k_4^2 + k_{12}^2 - k_3^2)
\nonumber\\
&&\pm \frac{1}{2 k_{12}^2}
\sqrt{4 k_1^2 k_{12}^2 - (k_1^2 + k_{12}^2 - k_2^2)^2}
\sqrt{4 k_4^2 k_{12}^2 \cos^2 \theta_4 -
(k_4^2 + k_{12}^2-k_3^2)^2}\,,
\label{k14_ito_123412costheta4}
\end{eqnarray}
which implies that the valid range of $\cos \theta_4$
is constrained by
\begin{eqnarray}
|\cos \theta_4| \geq \frac{|k_4^2 + k_{12} ^2 -k_3 ^2|}
{2 k_{12} k_4}\,.
\label{const_costheta4}
\end{eqnarray}

Motivated by the relation between the CMB trispectrum and the trispectrum for $\zeta$,
the shape function for the reduced trispectrum is defined as
\begin{eqnarray}
S_{\mathcal{T}}(k_1, k_2, k_3, k_4, k_{12}, \theta_4)
= (k_1 k_2 k_3 k_4)^2 k_{12}
\mathcal{T}_\zeta (k_1, k_2, k_3, k_4; k_{12}, \theta_4)\,.
\label{shapefunc_def}
\end{eqnarray}
Regan et al. \cite{Regan:2010cn} proposed to define an overlap between
two different shape functions $S_{\mathcal{T}}$ 
and $S_{\mathcal{T}'}$ as
\begin{eqnarray}
F(S_{\mathcal{T}}, S_{\mathcal{T}'})
= \int d \mathcal{V}_k \int d (\cos \theta_4)
S_{\mathcal{T}} (k_1, k_2, k_3, k_4, k_{12}, \theta_4)
S_{\mathcal{T}'} (k_1, k_2, k_3, k_4, k_{12}, \theta_4)
w(k_1, k_2, k_3, k_4, k_{12})\,,
\label{pritrispectrum_correlator_part}
\end{eqnarray}
where $w$ is an appropriate weight function.
The weight function should be chosen
such that $S^2 w$ in $k$ space produces the same scaling
as the estimator in $l$ space
and we adopt the one used in Ref.~\cite{Regan:2010cn},
\begin{eqnarray}
w (k_1, k_2, k_3, k_4, k_{12})
= \frac{1}{k_{12} (k_1 + k_2 + k_{12})(k_3 + k_4 + k_{12})}\,.
\label{trispectrum_weight}
\end{eqnarray}
With this choice of weight, the shape correlator is defined as
\begin{eqnarray}
\bar{\mathcal{C}} (S_{\mathcal{T}}, S_{\mathcal{T}'})
= \frac{F( S_{\mathcal{T}}, S_{\mathcal{T}'} )}
{\sqrt{F( S_{\mathcal{T}}, S_{\mathcal{T}} )
F( S_{\mathcal{T}'}, S_{\mathcal{T}'} ) }}\,.
\label{pritrispectrum_correlator}
\end{eqnarray}

\subsection{Parameterisation of six parameters}
First to parameterise the magnitude of the momenta,
we use the semiperimeter of the triangle
formed by the vectors ${\bf k}_1, {\bf k}_2, \k_1+\k_2$,
\begin{eqnarray}
q \equiv \frac12 (k_1 + k_2 + k_{12})\,.
\end{eqnarray}

From the scaling behaviour, the form of the shape
function on a constant-$q$ cross section becomes
independent of $q$ and we can write
\begin{eqnarray}
S_{\mathcal{T}} (k_1, k_2, k_3, k_4, k_{12},\theta_4)
= f(q) \bar{S}_{\mathcal{T}}
(\hat{k}_1, \hat{k}_2, \hat{k}_3, \hat{k}_4, \hat{k}_{12}, \theta_4)\,,
\label{pritrishapefunction}
\end{eqnarray}
where $\hat{k}_i \equiv k_i/q$ and $\hat{k}_{12}= k_{12}/q$.
Since we are restricted to the region where
the momenta $(k_1, k_2, k_{12})$ and $(k_3, k_4, k_{12})$
form triangles by momentum conservation, we will reparameterise
the allowed region to separate out the overall scale $q$
from the behaviour on a constant $q$ cross-sectional
slice. This five-dimensional slice is spanned by the
remaining coordinates. For triangle $(k_1, k_2, k_{12})$
we have
\begin{eqnarray}
k_{12} &=& q (1-\beta)\,,\\
k_1 &=& \frac{q}{2} (1 + \alpha + \beta)\,,\\
k_2 &=& \frac{q}{2} (1-\alpha + \beta)\,,
\end{eqnarray}
while for triangle  $(k_3, k_4, k_{12})$
\begin{eqnarray}
k_{12} &=& \epsilon q (1-\delta)\,,\\
k_3 &=& \frac{\epsilon q}{2}(1+\gamma + \delta)\,,\\
k_4 &=& \frac{\epsilon q}{2} (1-\gamma + \delta )\,,
\end{eqnarray}
where $\epsilon$ parameterises the ratio of the
perimeters of the two triangles, i.e.
$\epsilon = (k_3 + k_4 + k_{12})/(k_1 + k_2 + k_{12})$.
We do not lose the generality to consider $1 \leq \epsilon
< \infty$. The different expressions for $k_{12}$
imply that
\begin{eqnarray}
1-\beta = \epsilon (1-\delta )\,,
\end{eqnarray}
from which $\delta $ is eliminated to give
\begin{eqnarray}
k_3 &=& \frac{q}{2} (-1 + \beta + (2 + \gamma ) \epsilon )
\,,\\
k_4 &=& \frac{q}{2} (-1 + \beta + (2 - \gamma ) \epsilon )
\,.
\end{eqnarray}

The conditions for triangle $(k_1, k_2, k_{12})$
that $0 \leq k_1,k_2,k_{12} \leq q$ imply that
$0 \leq \beta \leq 1$ and $-(1-\beta) \leq \alpha
\leq 1- \beta $, while
the condition for triangle $(k_3, k_4, k_{12})$
that $0 \leq k_3, k_4, k_{12} \leq \epsilon k_{12}$
imply that $-(1-\beta)/\epsilon \leq \gamma \leq
(1-\beta)/ \epsilon$.
Furthermore, in terms of these variables,
the condition (\ref{const_costheta4}) is expressed as
\begin{eqnarray}
|\cos \theta_4| \geq \frac{|1+\beta^2
- \gamma \epsilon (-1+2 \epsilon) -\beta
(2 + \gamma \epsilon)|}{(1-\beta)
(-1+\beta+(2-\gamma) \epsilon)}\,.
\label{const_costheta4_ito_abce}
\end{eqnarray}
In summary, we have the following domains,
\begin{eqnarray}
0 \leq q < \infty,\;\; 1 \leq \epsilon < \infty\,,\;\;
0 \leq \beta \leq 1\,, \;\; -(1-\beta) \leq \alpha
\leq 1-\beta, \;\;-\frac{1-\beta }{\epsilon }
\leq \gamma \leq \frac{1-\beta }{\epsilon }\,,
\end{eqnarray}
together with Eq.~(\ref{const_costheta4_ito_abce}).
In practice, we introduce a cutoff for the integration
of $\epsilon$ as $S_{\mathcal{T}} S_{\mathcal{T}} w$ is
decreasing with $\epsilon$ asymptotically after integrating out
the dependence of $\alpha$, $\gamma$ and $\beta$ in the overlap integral.
we set the cut-off to be $\epsilon=10$ but the dependence on this
cut-off is very weak. Also we should emphasize that
the CMB measurements will never prove
the parameter region where $\epsilon \gg 1$.

Making use of this parameterisation, the shape function
(\ref{pritrishapefunction}), the weight function
(\ref{trispectrum_weight}) and the volume element
can be rewritten as
\begin{eqnarray}
S_{\mathcal{T}} (k_1, k_2, k_3, k_4, k_{12},\theta_4)
&=& f(q) \bar{S}_{\mathcal{T}} (\alpha,\beta,\gamma,\epsilon,\theta_4)
\,,\nonumber\\
w(k_1, k_2, k_3, k_4, k_{12}) &=& \frac{1}{4 \epsilon
(1-\beta)}\,,\\
d \mathcal{V}_k = d k_1 dk_2 dk_3 dk_4 dK
&=& \epsilon q^4 dq d\alpha d\beta d\gamma d\epsilon\,.
\end{eqnarray}
It is worth noting that although the integration
$d \mathcal{V}_k$ is five-dimensional
in Eq.~(\ref{pritrispectrum_correlator_part}),
for scale-invariant shape functions with constant
$f (q)$, it is enough to evaluate shape correlations
only for the four dimensional slices with constant $q$.

\section{Shape correlations}
In this section, we study the overlap between Eq.~(\ref{fiducial}) and
the trispectra in single field and multi-field DBI inflation.
It is worth mentioning that from the definition of
the shape correlator (\ref{pritrispectrum_correlator}),
the shape correlations are independent of
the normalisations of shape functions. We will discuss the normalisation
of the trispectrum in the next section.

\subsection{Equilateral shape}
First, we find that the shape function for the trispectrum
(\ref{fiducial}) is given by
\begin{eqnarray}
S_{\mathcal{T}} ^{equil} &=& N^{equil}
S_{\mathcal{T}} ^{c_1}\,,
\label{rel_shape_equil_shape_c1}\\
N^{equil} &=&
\frac{64}{3} {\cal P}_\zeta^3 g_{\rm NL} ^{equil}
\,,\label{norm_equil}
\end{eqnarray}
where $S_{\mathcal{T}} ^{c_1}$ is given by
Eq.~(\ref{shape_c1}). We will assume the scale independence of the spectrum
${\cal P}_{\zeta}$ in the rest of the paper.
In k-inflation model, this class of models are characterised
by $P_{,4X} \gg X^{-2} P_{,XX}, X^{-1} P_{,XXX}$ for the action given by Eq.~(\ref{action}).
It was also shown that, in the context of the effective theory of inflation,
it is possible to construct consistent inflationary models where the trispectrum
is characterised by this shape function and its amplitude is not constrained by
the bispectrum \cite{Senatore:2010jy}.

As this shape function depends on $\alpha$, $\beta$, $\gamma$, $\epsilon$,
first, we clarify the $\epsilon$ dependence of the signal
which is given by $S_{\mathcal{T}} ^{equil} S_{\mathcal{T}} ^{equil} w$.
We find that after integrating out the dependence of
$\alpha$, $\beta$ and $\gamma$ in the overlap integration,
the amplitude of the signal is proportional
to $1/\epsilon^7$ asymptotically. This shows that the dominant contribution to
the signal for this shape is coming from $\epsilon \sim 1$.
In Fig.~1, we show the $\epsilon$ dependence of $S_{\mathcal{T}} ^{equil} S_{\mathcal{T}} ^{equil} w$.

\begin{figure}[t]
\scalebox{1}{
\centerline{
\includegraphics{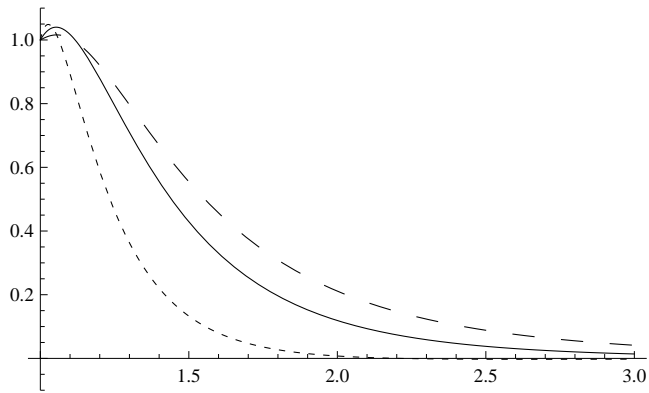}}}
\caption{
We show the $\epsilon$ dependence of $S_{\mathcal{T}} ^{equil}
S_{\mathcal{T}} ^{equil} w $ (solid line),
$S_{\mathcal{T}} ^{DBI(\sigma)}
S_{\mathcal{T}} ^{equil} w $ (dashed line)
and $S_{\mathcal{T}} ^{DBI(s)}
S_{\mathcal{T}} ^{equil} w $ (dotted line)
by integrating out the dependence of
$\alpha$, $\beta$ and $\gamma$ in the overlap integration.
We find that $S_{\mathcal{T}} ^{DBI(\sigma)}
S_{\mathcal{T}} ^{equil} w $
and $S_{\mathcal{T}} ^{DBI(s)}
S_{\mathcal{T}} ^{equil} w $
asymptote to $\propto 1/\epsilon^4$ while
$S_{\mathcal{T}} ^{equil}
S_{\mathcal{T}} ^{equil} w $ $\propto 1/\epsilon^7$ for large $\epsilon$.
We find that while
$S_{\mathcal{T}} ^{DBI(\sigma)}
S_{\mathcal{T}} ^{equil} w $ is always positive,
$S_{\mathcal{T}} ^{DBI(s)}
S_{\mathcal{T}} ^{equil} w $ become negative
above some critical value of $\epsilon$.
Because of this, the full overlap between
$S^{DBI(s)} _{\mathcal{T}}$ and $S^{equil} _{\mathcal{T}}$
takes much smaller value than the one estimated at
configurations with $\epsilon=1$. We have normalised
so that the values become $1$ at $\epsilon=1$.}
\label{epsilon_dep2}
\end{figure}

\begin{figure}[h]
\scalebox{0.7}{
\centerline{
\includegraphics{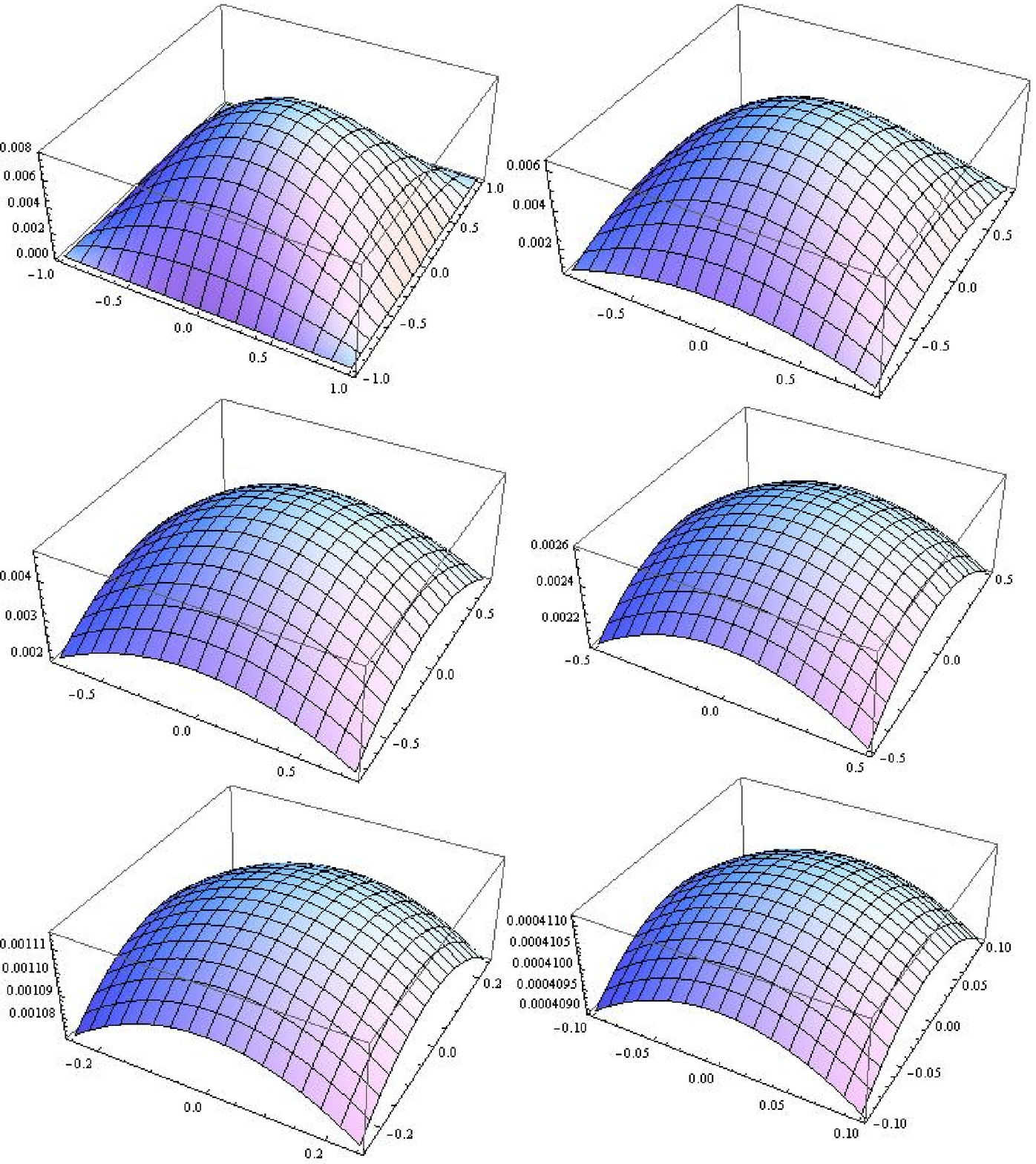}}}
\caption{
In this group of figures, we plot
$S_{\mathcal{T}} ^{equil} (\alpha, \beta, \gamma, \epsilon)$
with $\epsilon=1$. $\beta=0, 0.1, 0.25, 0.5, 0.75, 0.9$
for upper left, upper right, middle left, middle right,
lower left, lower right, respectively.
Vertical axes for each plot denote $\alpha$ and $\gamma$.
For these configurations, $S_{\mathcal{T}} ^{equil}$
is symmetric under the exchange of $\alpha$ and $\gamma$ and
the physical region is given by $1-\beta > \alpha, \gamma
> \beta-1$.
We set $N^{equil} =1$.}
\label{c1_alpha_gamma_beta}
\end{figure}

Next, we examine the $\alpha$, $\beta$, $\gamma$
dependence of $S_\mathcal{T} ^{equil}$.
For this purpose, we plot
$S_\mathcal{T} ^{equil} (\alpha, \beta, \gamma, \epsilon)$
evaluated at $\epsilon=1$ for given $\beta$
in Fig.~{\ref{c1_alpha_gamma_beta}} where
$S_{\mathcal{T}} ^{equil}$ is symmetric under the exchange of
$\alpha$ and $\gamma$ for the configurations with $\epsilon =1$ and
the physical region is given by $1-\beta > \alpha, \gamma
> \beta-1$.

\subsection{Single filed DBI inflation}

It was suggested in \cite{Arroja:2009pd} that the shape function corresponding to
the reduced trispectrum of single field DBI inflation 
at leading order in
the slow-roll expansion is given by
\begin{eqnarray}
S^{DBI(\sigma)} _{\mathcal{T}}
&=&N^{DBI(\sigma)}\left[
-3 S^{c_{1}} _{\mathcal{T}} +
\frac{1}{64} S^{s_{1}} _{\mathcal{T}}
+\frac{1}{64} S^{s_{2}} _{\mathcal{T}}
-\frac{1}{64} S^{s_{3}} _{\mathcal{T}}
\right]\,,
\label{dbisigma_temp}\\
N^{DBI(\sigma)}&=& \frac{H^{12}}{\dot{\phi}^6 c_s^4}\,,
\label{norm_sdbi}
\end{eqnarray}
where  $S^{s_{1}} _{\mathcal{T}}$,
$S^{s_{2}} _{\mathcal{T}}$ and
$S^{s_{3}} _{\mathcal{T}}$
are given by Eqs.~(\ref{shape_s1}),
(\ref{def_shape_s2})
and (\ref{def_shape_s3}), respectively.

Similar to the case of the equilateral shape, first
we examine the $\epsilon$ dependence. In Fig.~1, we plot
$S_{\mathcal{T}} ^{DBI(\sigma)} S_{\mathcal{T}} ^{equil} w$.
We find that after integrating out the dependence of
$\alpha$, $\beta$ and $\gamma$ in the overlap integration,
the amplitude of the signal is proportional
to $1/\epsilon^4$ asymptotically. This shows that the dominant contribution to the
overlap between the single field DBI model and the equilateral shape is
coming from $\epsilon \sim 1$. The difference of the asymptotic $\epsilon$
dependence between the trispectra corresponding to the differences of the shapes
between $S_{\mathcal{T}} ^{c_1}$, which is coming from the contact interaction and
$S_{\mathcal{T}} ^{s_1}$, $S_{\mathcal{T}} ^{s_2}$,
$S_{\mathcal{T}} ^{s_3}$, which arise from the scalar exchanges.
This was pointed out in Ref.~\cite{Chen:2009bc} by considering
the double squeezed limit ($k_3 = k_4=k_{12} \to 0$).
Therefore, it is natural that the asymptotic $\epsilon$ dependence between
$S_{\mathcal{T}} ^{equil}$ and $S_{\mathcal{T}} ^{DBI(\sigma)}$ is different,
as $S_{\mathcal{T}} ^{DBI(\sigma)}$ is obtained by a linear combination of
$S_{\mathcal{T}} ^{s_1}$, $S_{\mathcal{T}} ^{s_2}$
and $S_{\mathcal{T}} ^{s_3}$.

\begin{figure}[h]
\scalebox{0.7}{
\centerline{
\includegraphics{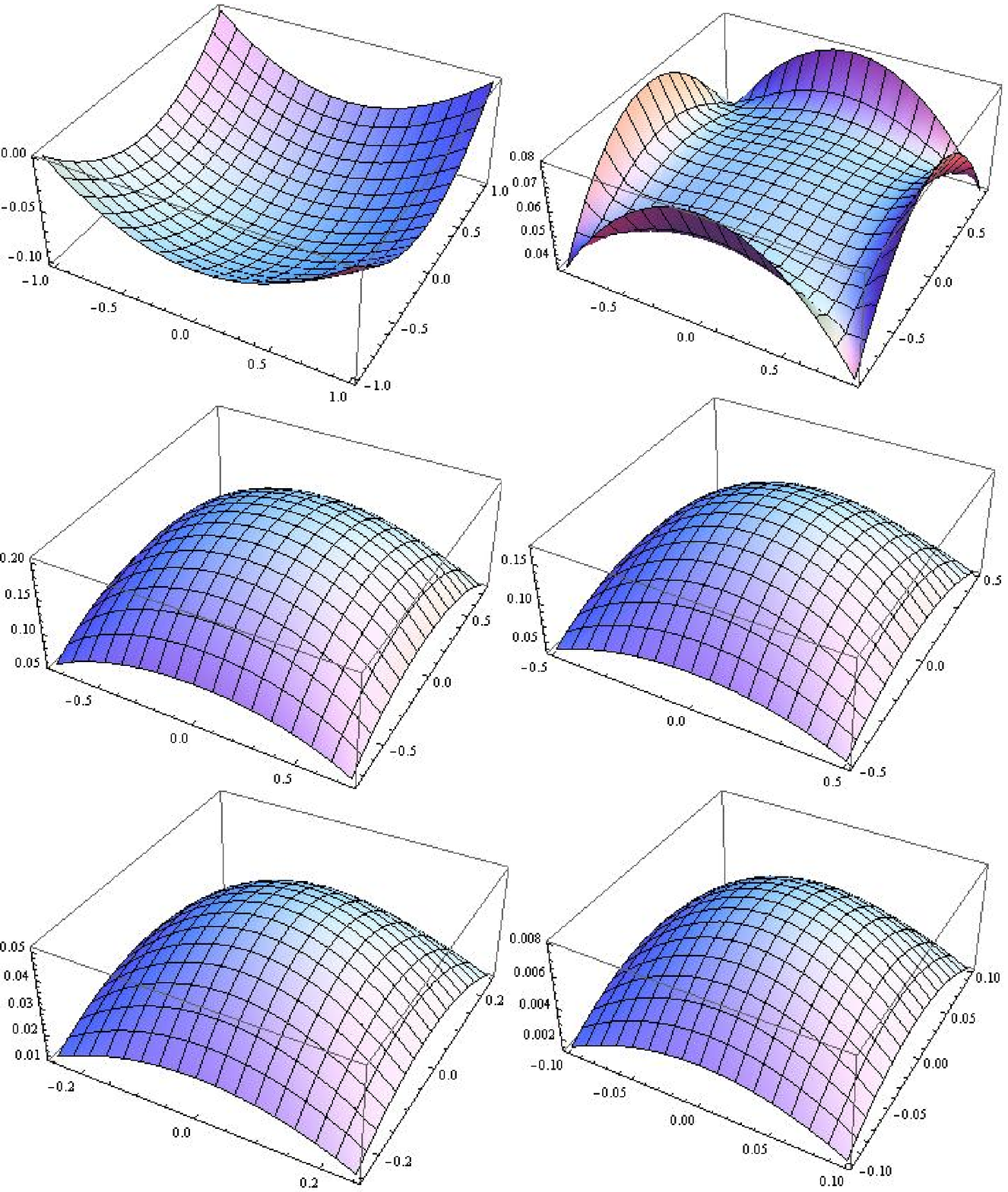}}}
\caption{
In this group of figures, we plot
$S_{\mathcal{T}} ^{DBI(\sigma)} (\alpha, \gamma, \beta)$
with $\epsilon=1$. $\beta=0, 0.1, 0.25, 0.5, 0.75, 0.9$
for upper left, upper right, middle left, middle right,
lower left, lower right, respectively.
Vertical axes for each plot denote $\alpha$ and $\gamma$.
For these configurations,
$S_{\mathcal{T}} ^{DBI(\sigma)}$
is symmetric under the exchange of
$\alpha$ and $\gamma$ and
the physical region is given by $1-\beta > \alpha, \gamma
> \beta-1$. We set $N^{DBI(\sigma)} =1$.
}
\label{sdbi_alpha_gamma_beta}
\end{figure}

Next, we examine the $\alpha$, $\beta$, $\gamma$
dependence of $S_{\mathcal{T}} ^{DBI(\sigma)}$.
For this purpose, we plot
$S_{\mathcal{T}} ^{DBI(\sigma)}
(\alpha, \beta, \gamma, \epsilon)$ evaluated
at $\epsilon=1$ for given $\beta$
in Fig.~{\ref{sdbi_alpha_gamma_beta}}
where
$S_{\mathcal{T}} ^{DBI(\sigma)}$ is symmetric under the exchange of
$\alpha$ and $\gamma$ for the configurations with $\epsilon =1$ and
the physical region is given by $1-\beta > \alpha, \gamma
> \beta-1$.
Except for the region with very small value of $\beta$
(from $0$ to $\sim 0.1$) the shapes are very similar
to $S_{\mathcal{T}} ^{equil}$. This explains that
the overlap between
$S_{\mathcal{T}} ^{DBI(\sigma)}$ and
$S_{\mathcal{T}} ^{equil}$ is sufficiently large
for the configurations with $\epsilon=1$.

Table~\ref{table_summary} provides a summary of correlations between
$S_\mathcal{T} ^{equil}$ and $S_{\mathcal{T}} ^{DBI(\sigma)}$.
In addition to the correlation considering full configurations
dealing with five dimensional parameter space,
for comparisons, we also consider the configurations
limited with $\epsilon=1$ and equilateral
configurations ($\epsilon=1$, $\alpha=\gamma=0$).
The overlap decrease if we include the non-equilateral configurations keeping $\epsilon=1$.
This is due to the difference of the shapes for small $\beta$. Also the different asymptotic
behaviours with respect to $\epsilon$ further reduces the overlap if we integrate over
$\epsilon$. However, even after we perform the all integration, the overlap remains
high at around 87$\%$ level.

\subsection{Multi-field DBI inflation}

It was suggested  in \cite{Mizuno:2009mv}
that the reduced trispectrum
of multi-field DBI inflation dominated by originally purely entropic
perturbations at leading order in the slow-roll expansion
is given by
\begin{eqnarray}
S^{DBI(s)}_{\mathcal{T}}  &=&
N^{DBI(s)}\left[-\frac{1}{8} S^{c_{2}} _{\mathcal{T}} +
\frac{1}{576} S^{s_{1}} _{\mathcal{T}}
+\frac{1}{64} S^{\tilde{s}_{2}} _{\mathcal{T}}
+\frac{1}{192} S^{\tilde{s}_{3}} _{\mathcal{T}}
\right]\,,
\label{dbis_temp}\\
N^{DBI(s)} &=& \frac{H^{12}}{\dot{\phi}^6 c_s^4}
T_{\mathcal{R} S}^4\,,
\label{norm_mdbi}
\end{eqnarray}
where $S^{\tilde{s}_{2}} _{\mathcal{T}}$ and
$S^{\tilde{s}_{3}} _{\mathcal{T}}$
are given by Eqs.~(\ref{def_shape_s2_tilde})
and (\ref{def_shape_s3_tilde}), respectively.

Again, we first examine the $\epsilon$ dependence of
the overlap (in Fig.~1).
We find that after integrating out the dependence of
$\alpha$, $\beta$ and $\gamma$,
the amplitude of the overlap, $S_{\mathcal{T}} ^{equil}
S_{\mathcal{T}} ^{DBI(s)} w$ is proportional
to $1/\epsilon^4$ asymptotically as in the single field case.
This shows that the dominant contribution to the signal
for this shape is coming from $\epsilon \sim 1$.

The asymptotic $\epsilon $ dependence of
$S_{\mathcal{T}} ^{DBI(s)}$ is the same as
that of $S_{\mathcal{T}} ^{DBI(\sigma)}$.
We find that asymptotically
$S_{\mathcal{T}} ^{s_{2b}}$, $S_{\mathcal{T}} ^{s_{2d}}$
and $S_{\mathcal{T}} ^{s_{3b}}$
given by Eqs.~(\ref{shape_s2b}), (\ref{shape_s2d})
and (\ref{shape_s3b}) give the dominant
contribution to both $S_{\mathcal{T}} ^{DBI(\sigma)}$
and $S_{\mathcal{T}} ^{DBI(s)}$,
which characterises the asymptotic $\epsilon$ dependence.
However, as is shown in Fig.~{\ref{epsilon_dep2}},
$S_{\mathcal{T}} ^{DBI(s)} S_{\mathcal{T}} ^{equil} w $ become
negative above some critical value of $\epsilon$. This reduces
the final overlap once we integrate over $\epsilon$.

\begin{figure}[h]
\scalebox{0.7}{
\centerline{
\includegraphics{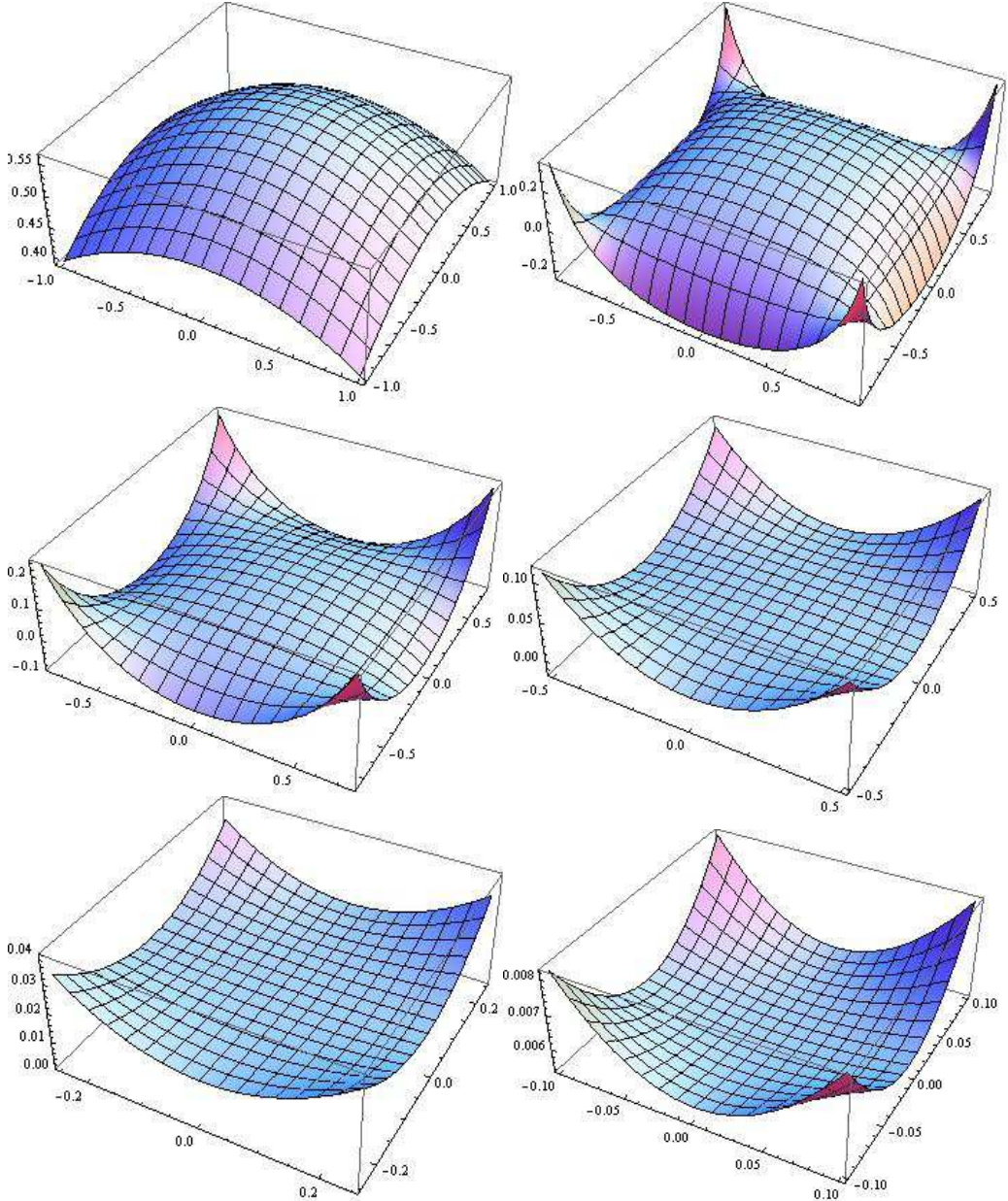}}}
\caption{
In this group of figures, we plot
$S_{\mathcal{T}} ^{DBI(s)} (\alpha, \gamma, \beta)$
with $\epsilon=1$. $\beta=0, 0.1, 0.25, 0.5, 0.75, 0.9$
for upper left, upper right, middle left, middle right,
lower left, lower right, respectively.
Vertical axes for each plot denote $\alpha$ and $\gamma$.
For these configurations,
$S_{\mathcal{T}} ^{DBI(s)}$
is symmetric under the exchange of
$\alpha$ and $\gamma$ and
the physical region is given by $1-\beta > \alpha, \gamma
> \beta-1$.
We set
$N^{DBI(s)} =1$.}
\label{mdbi_alpha_gamma_beta}
\end{figure}

Next, we examine  the $\alpha$, $\beta$, $\gamma$
dependence of $S_{\mathcal{T}} ^{DBI(s)}$.
For this purpose, again we plot
$S_{\mathcal{T}} ^{DBI(s)}
(\alpha, \beta, \gamma, \epsilon)$ evaluated
at $\epsilon=1$ for given $\beta$
in Fig.~{\ref{mdbi_alpha_gamma_beta}}
where
$S_{\mathcal{T}} ^{DBI(s)}$ is symmetric
under the exchange of
$\alpha$ and $\gamma$ for the configurations with
$\epsilon =1$ and
the physical region is given by $1-\beta > \alpha, \gamma
> \beta-1$.

Except for the region with very small value of $\beta$
(from $0$ to $\sim 0.1$) the shape is different from
$S_{\mathcal{T}} ^{equil}$.
This explains that the overlap between
$S_{\mathcal{T}} ^{DBI(s)}$ and
$S_{\mathcal{T}} ^{equil}$ is not so large
even for the configurations with $\epsilon=1$.

Table~\ref{table_summary}
provides a summary of correlations between
$S_\mathcal{T} ^{equil}$ and
$S_{\mathcal{T}} ^{DBI(s)}$.
As in the single field case, in addition to the correlation considering full configurations
dealing with five dimensional parameter space,
for comparisons, we also consider the configurations
limited with $\epsilon=1$ and equilateral
configurations ($\epsilon=1$,
$\alpha=\gamma=0$).

Table~\ref{table_summary} shows that the overlap becomes
smaller once we include the non-equilateral configurations
with $\epsilon=1$. This is clear from the shape difference for
$\beta >0.1$. In addition the shape correlation
for full configurations becomes further smaller once
we integrate over $\epsilon$. As explained before,
this is due to the fact that
while $S_{\mathcal{T}} ^{DBI(\sigma)}
S_{\mathcal{T}} ^{equil} w $ is always positive,
$S_{\mathcal{T}} ^{DBI(s)}
S_{\mathcal{T}} ^{equil} w $ become negative
above some critical value of $\epsilon$.
This confirms the fact that the shape dependence of
trispectrum can in principle distinguish multi-field DBI inflation
models form single field DBI inflation models
shown by Refs.~\cite{Mizuno:2009cv, Mizuno:2009mv}.
In practice, the overlap still remains at 33$\%$ level after integrating
over all the shape parameters and the equilateral
shape could still be used to get a reasonable estimation for the constraints on
multi-field DBI inflation model.

\section{Theoretical predictions for $g_{\rm NL}^{equil}$}

In this section, making use of the shape correlations
investigated in the previous section,
we give theoretical predictions for $g_{\rm NL} ^{equil}$.
As a consistency check, we applied the same method to
estimate the amplitude of the bispectrum in DBI inflation
in Appendix~\ref{sec_bispectrum_estimation}.

In k-inflation models, as is shown in Appendix~{\ref{parts}}
by setting $P_{,4X} \gg X^{-2} P_{,XXX}, X^{-1} P_{,XXX}$,
the shape function is given by
\begin{eqnarray}
S^{equil} _{\mathcal{T}}
= \frac{H^{12} P_{,4X}}{16 P_{,X}^4 c_s}
S^{c_1} _{\mathcal{T}}\,.
\label{shape_func_equi_effect}
\end{eqnarray}
Then, by comparing
Eqs.~(\ref{rel_shape_equil_shape_c1}) with
(\ref{shape_func_equi_effect}),
we find $g_{\rm NL} ^{equil}$ is obtained as
\begin{eqnarray}
g_{\rm NL} ^{equil} =\frac{3 X^3 c_s^2 P_{, 4X}}
{16 P_{,X}}\,,
\label{gnl_eff_model}
\end{eqnarray}
where we have used ${\cal P}_\zeta  = H^4/ (4 X c_s P_{,X} )$
for single field k-inflation.

In order to express the amplitude of trispectrum
in single field DBI inflation in terms of
$g_{\rm NL} ^{equil}$, we rewrite
Eq.~(\ref{dbisigma_temp}) in the following form:
\begin{eqnarray}
S^{DBI(\sigma)} _{\mathcal{T}}
&=& 0.41 {\cal P}_\zeta^3 \frac{g_{\rm NL} ^{equil}}
{\bar{\mathcal{C}}(S^{DBI(\sigma)}_{\mathcal{T}},
S^{equil} _{\mathcal{T}})}\left[
-3 S^{c_{1}} _{\mathcal{T}} +
\frac{1}{64} S^{s_{1}} _{\mathcal{T}}
+\frac{1}{64} S^{s_{2}} _{\mathcal{T}}
-\frac{1}{64} S^{s_{3}} _{\mathcal{T}}
\right]\,,
\label{shape_func_sdbi_est}
\end{eqnarray}
where the numerical factor
in Eq.~(\ref{shape_func_sdbi_est}) is chosen so that when
$g_{\rm NL} ^{equil}=1$ and
$\bar{\mathcal{C}}
(S^{DBI(\sigma)} _{\mathcal{T}}  ,
S^{equil} _{\mathcal{T}}  ) =1$,
the following conditions are satisfied,
\begin{eqnarray}
F (S^{DBI(\sigma)} _{\mathcal{T}},
S^{DBI(\sigma)} _{\mathcal{T}})
&=& F(S^{equil} _{\mathcal{T}},
S^{equil} _{\mathcal{T}})\,,
\nonumber\\
F(S^{DBI(\sigma)} _{\mathcal{T}},
S^{equil} _{\mathcal{T}}) &>&0\,.
\label{def_sdbi_uni}
\end{eqnarray}
Of course,
$\bar{\mathcal{C}}
(S^{DBI(\sigma)} _{\mathcal{T}}  ,
S^{equil} _{\mathcal{T}}  ) =1$
is not true in reality and this factor will
enhance the amplitude of the signal
for a given $g_{\rm NL} ^{equil}$.
This term is necessary because when we use the estimator related with
$S^{equil} _{\mathcal{T}}$ for the signal whose shape is
characterised by $S^{DBI(\sigma)} _{\mathcal{T}} $,
the observed signal is suppressed by
 $\bar{\mathcal{C}}
(S^{DBI(\sigma)} _{\mathcal{T}}  ,
S^{equil} _{\mathcal{T}}  )$
and it is necessary to compensate this.
Then, by comparing Eqs.~(\ref{dbisigma_temp})
with (\ref{def_sdbi_uni}), we can relate
 $g_{\rm NL} ^{equil}$ with the sound speed $c_s^2$
as
\begin{eqnarray}
g_{NL} ^{equil} = \frac{17}{c_s^4}\,,
\label{gnl_sdbi}
\end{eqnarray}
where we have used
${\cal P}_\Phi = H^4/ (2 \dot{\phi}^2)$
for single field DBI inflation.

Similarly, in order to express the amplitude of trispectrum
in multi-field DBI inflation in terms of
$g_{\rm NL} ^{equil}$, we rewrite
Eq.~(\ref{dbis_temp}) in the following form:
\begin{eqnarray}
S^{DBI(s)}_{\mathcal{T}}  &=&
1.2 {\cal P}_\zeta^3
\frac{g_{\rm NL} ^{equil}}
{\bar{\mathcal{C}}(S^{DBI(s)}_{\mathcal{T}},
S^{equil} _{\mathcal{T}})}
\left[-\frac{1}{8} S^{c_{2}} _{\mathcal{T}} +
\frac{1}{576} S^{s_{1}} _{\mathcal{T}}
+\frac{1}{64} S^{\tilde{s}_{2}} _{\mathcal{T}}
+\frac{1}{192} S^{\tilde{s}_{3}} _{\mathcal{T}}
\right]\,,
\label{shape_func_mdbi_est}
\end{eqnarray}
where the numerical factor
in Eq.~(\ref{shape_func_mdbi_est}) is again
chosen so that when
$g_{\rm NL} ^{equil}=1$ and
$\bar{\mathcal{C}}
(S^{DBI(s)} _{\mathcal{T}}  ,
S^{equil} _{\mathcal{T}}  ) =1$,
the following conditions are satisfied,
\begin{eqnarray}
F (S^{DBI(s)} _{\mathcal{T}},
S^{DBI(s)} _{\mathcal{T}})
&=& F(S^{equil} _{\mathcal{T}},
S^{equil} _{\mathcal{T}})\,,
\nonumber\\
F(S^{DBI(s)} _{\mathcal{T}},
S^{equil} _{\mathcal{T}}) &>&0\,.
\label{def_mdbi_uni}
\end{eqnarray}

Then, by comparing Eqs.~(\ref{dbis_temp})
with (\ref{def_mdbi_uni}), we can relate
 $g_{\rm NL} ^{equil}$ with the sound speed $c_s^2$ and the transfer
coefficient that relate the amplitude of original entropy perturbations
to the final curvature perturbation $T_{\mathcal{R} S}^2$
as
\begin{eqnarray}
g_{NL} ^{equil} = \frac{2.2}{c_s^4 T_{\mathcal{R} S}^2}\,,
\label{gnl_mdbi}
\end{eqnarray}
where we have used ${\cal P}_\zeta = H^4 T_{\mathcal{R} S}^2
/(2 \dot{\phi}^2)$ for multi field DBI inflation.

In Table~\ref{table_summary},
we summarise theoretical predictions for
$g^{equil} _{\rm NL}$ for the models
discussed in this section.

\begin{widetext}
\begin{center}
\begin{table} [h!]
\begin{tabular} { |c|c|c|c|c|c| }
\hline
  &Overlap-full & $\epsilon=1$ & equilateral&
 theoretical prediction for $g^{equil} _{\rm NL}$ &
 $f^{equil}_{\rm NL}$ \\
 \hline equilateral shape
&  $1$ & $1$  & $1$& $(3 X^3 c_s^2 P_{, 4X})/(16 P_{,X})$
& $f^{equil}_{\rm NL}$ \\
\hline
single DBI
&  $0.87$ & $0.90$  & $0.92$&
$17 / c_s^4$  & $-0.36/c_s^2$
 \\
\hline
multi DBI
&  $0.33$ & $0.60$  & $0.85$&
$2.2 / (c_s^4 T_{\mathcal{R} S}^2)$ &$ -0.36/(c_s^2 T_{\mathcal{R} S}^2)$
 \\
\hline
\end{tabular}
\caption{Shape correlations
against
$S_\mathcal{T} ^{equil}$
for full
configurations, the configurations restricted to
$\epsilon=1$, equilateral configurations
($\epsilon=1$, $\alpha=\gamma=0$)
in the model with equilateral shape
motivated by effective theory of inflation,
single field DBI inflation and multi-field DBI inflation.
We also summarise theoretical predictions for
$g^{equil} _{\rm NL}$ and $f_{\rm NL}^{equil}$ in these models.}
\label{table_summary}
\end{table}
\end{center}
\end{widetext}

It is instructive to compare the values (\ref{gnl_sdbi})
and (\ref{gnl_mdbi}) with previous results of
$\tau_{\rm NL}$ \cite{Mizuno:2009mv}
based on the matching of the amplitude at
a specific equilateral configuration. We define the non-linear parameter
\begin{equation}
\tau_{\rm NL} = \frac{T_{\zeta}(\k_1,\k_2,\k_3,\k_4) k_1^3 k_2^3 k_3^3 k_4^3
{\cal P}_{\zeta}^{-3}}{
[(k_1^3 k_2^3 + k_3^3 k_4^3)(k_{13}^{-3} + k_{14}^{-3})
+ (k_1^3 k_4^3 + k_2^3 k_3^3)(k_{12}^{-3} + k_{13}^{-3})
+ (k_1^3 k_3^3 + k_2^3 k_4^3)(k_{12}^{-3} + k_{14}^{-3}]},
\end{equation}
and evaluated it for the configuration specified by
$k_1=k_2=k_3=k_4=k$, $k_{12}=k_{13} = k_{14} = (2/\sqrt{3})k$.
For single field and multi-field DBI inflation models, we obtain
\begin{eqnarray}
\tau_{\rm NL} &=& \frac{0.56}{c_s^4}\,,\;\;
{\rm for \;\; single\;\;field\;\;DBI\;\;inflation}
\label{tausDBI}\\
\tau_{\rm NL} &=& \frac{0.12}{c_s^4 T_{\mathcal{R} S}^2}\,,
\;\;{\rm for \;\; multifield\;\;DBI\;\;inflation}
\label{taumDBI}
\end{eqnarray}
Using the same procedure, we get $\tau_{\rm NL} =
2 g_{\rm NL} ^{equil}/9 \sqrt{3}$. By comparing this to
(\ref{tausDBI}) and (\ref{taumDBI})
we can estimate $g_{\rm NL}^{equil}$ as
\begin{eqnarray}
g_{\rm NL} ^{equil} &=& \frac{4.4}{c_s^4}\,,\;\;
{\rm for \;\; single\;\;field\;\;DBI\;\;inflation}
\label{gnl_sdbi_prev}\\
g_{\rm NL} ^{equil}&=& \frac{0.94}
{c_s^4 T_{\mathcal{R} S}^2}\,,
\;\;{\rm for \;\; multifield\;\;DBI\;\;inflation}
\label{gnl_mdbi_prev}
\end{eqnarray}
which underestimates the amplitude by factor $2 \sim 4$
compared with the results of Eqs.~(\ref{gnl_sdbi}) and (\ref{gnl_mdbi}).
This demonstrates that unlike the bispectrum case where the matching
of the amplitude at the equilateral configuration gives a reasonable estimation
for $f_{\rm NL}^{equil}$, it is necessary to calculate the
overlap between the shapes in full five-dimensional parameter space to extract
the amplitude of the trispectrum $g_{\rm NL}^{equil}$.

\section{Conclusion}

It is well known that there are many interesting
early universe models that predict equilateral type
primordial non-Gaussianity motivated by string theory
and effective field theory.
Taking into account the fact that
future experiments such as Planck can prove even next order
statistics, it is important to study the
primordial trispectrum in these models.
For example, we had shown previously that the trispectrum can
in principle distinguish multi-field DBI inflation models
from single field DBI inflation models from
the shape dependence
\cite{Arroja:2009pd, Mizuno:2009cv, Mizuno:2009mv}.
On the other hand, at the practical level,
since the form of the trispectrum is too complicated,
the estimator in this class of models had not been implemented
explicitly.

Therefore, in this work we have presented a method to estimate
primordial trispectrum in equilateral type non-Gaussian models
such as k-inflation model whose action is given by Eq.~(\ref{action}),
single field DBI inflation and multi-field DBI inflation. Our method is based on
the following two facts. One is that the equilateral shape
given by Eq.~(\ref{fiducial}) becomes factorisable
by introducing the integral $1/M^n =(1/\Gamma(n))
\int^{\infty}_{0} t^{(n-1)} e^{-M t}$ as was suggested
in Ref.~\cite{Chen:2009bc}. The other is that in terms of
the shape correlation proposed by Ref.~\cite{Regan:2010cn},
we can relate the amplitudes of trispectra with different
shapes.

After reviewing the shape correlator, we have calculated
the overlaps between the equilateral shape
and the shapes of trispectra in single field DBI inflation and
multi-field DBI inflation. We have shown that
the shape is $87 \%$ correlated with
the one in single field DBI inflation, while it is
$33 \%$ correlated with that in
multi-field DBI inflation when the curvature perturbation
is originated from purely entropic perturbations during inflation.
We have summarised the overlaps including
the configurations restricted to $\epsilon=1$ and
equilateral configurations ($\epsilon=1$, $\alpha=\gamma=0$)
in Table~\ref{table_summary}.
We found that the main difference between the
the equilateral shape and the shape in single field
DBI inflation comes from the configurations with
$\beta \sim 0$,
which can be seen even
in the equilateral configurations ($k_1=k_2=k_3=k_4$).
For the shape in multi-field DBI inflation, as the behaviour
of the shape function is different from the one in
single field DBI inflation
\cite{Arroja:2009pd, Mizuno:2009cv, Mizuno:2009mv},
the overlap becomes smaller. Regardless of this,
when we take into account of the fact that this overlap
is calculated in the five-dimensional parameter space, the
$33 \%$ correlation is not necessarily small.
For example, the overlap between equilateral shape and local
shapes, which depend on cutoffs in the integration due to divergences
in various limits, is less than $2 \%$.

Then, we have given theoretical predictions
for $g_{\rm NL} ^{equil}$, which enables us to
constrain this type of non-Gaussian models from future experiments.
For the model with equilateral shape motivated by
k-inflation, we obtained
$g_{\rm NL} ^{equil} = (3 X^3 c_s^2 P_{,4X})/(16 P_{,X})$,
while for single field DBI inflation and
multi-field DBI inflation, $g_{\rm NL} ^{equil} = 17/c_s^4$
and $g_{\rm NL} ^{equil}=
2.2/(c_s^4 T_{\mathcal{R} S}^2)$, respectively.
To obtain this value, instead of matching
the amplitudes of the shape functions at a specific point
in the parameter space, we have adopted an overlap function,
$F(S_{\mathcal{T}}, S_{\mathcal{T}'})$, defined in
Eq.~(\ref{pritrispectrum_correlator_part}), which involves
integration over five-parameters.

Before closing, let us comment on the detectability
of the trispectrum in future experiments.
According to the estimation in Refs.~\cite{Creminelli:2006gc, Senatore:2010jy},
the observational errors on $f_{\rm NL}$ and
$g_{\rm NL}$ scales as
\begin{eqnarray}
\Delta f_{\rm NL} \sim \frac{1}
{{\cal P}_\zeta^{1/2} N_{\rm pix} ^{1/2}}\,,
\;\;\;\; \Delta g_{\rm NL} \sim \frac{1}
{{\cal P}_\zeta N_{\rm pix} ^{1/2}}\,,
\label{obs_const_errors}
\end{eqnarray}
where $N_{\rm pix}$ represents the number of data points
of the experiment. Therefore, current limit on $g_{\rm NL}$
by WMAP is of  order $10^7$, while the future experiments
like Planck \cite{Planck} and 21-cm line experiments
\cite{Loeb:2003ya} are expected to produce a limit
$g_{\rm NL} \sim 10^6$ and  $g_{\rm NL} \sim 10^3$,
respectively.

For the models like single field DBI inflation
where non-Gaussian parameters are given by
$f_{\rm NL} \sim c_s^{-2}$ and $g_{\rm NL} \sim c_s^{-4}$,
the trispectrum is not detectable even by the Planck satellite
since there is already a constraint like $c_s^2 \geq 10^{-2}$
from $f_{\rm NL} \leq 10^2$ from the bispectrum measurement \cite{Senatore:2010jy}.
However, for multi-field DBI inflation models
where non-Gaussian
parameters are given by $f_{\rm NL} \sim c_s^{-2} T_{\mathcal{R}S}^{-2}$
and $g_{\rm NL} \sim c_s^{-4} T_{\mathcal{R}S}^{-2}$,
it might be possible to detect the trispectrum if there is a large transfer from
the entropy mode. For example, it is detectable by Planck if
$T_{\mathcal{R}S}=10$. In this context, to construct a concrete
theoretical model which gives a large transfer coefficient
$ T_{\mathcal{R}S}$ is important.
The model with the equilateral shape motivated by
effective theory of inflation \cite{Senatore:2010jy}
can give $g_{\rm NL}$ much larger than $10^6$
which is detectable by Planck while keeping the value of
$f_{\rm NL}$ to be just of order one.

Finally, although we have not studied in this paper,
it is known that the ghost inflation also gives
equilateral type non-Gaussianity
\cite{ArkaniHamed:2003uz, Senatore:2004rj} and recently
the shape dependence of the trispectrum was also calculated
\cite{Izumi:2010wm, Huang:2010ab}. It might be interesting
to express the amplitude of the trispectrum
in terms of the estimator proposed in this paper.

\begin{acknowledgments}
We would like to thank Rob Crittenden and Dominic Galliano for useful discussions. 
KK thanks the Yukawa Institute for Theoretical Physics, Kyoto University and the Royal Society for two workshops, ``Non-linear cosmological perturbations'' (YITP-W-09-01) and ``The non-Gaussian universe'' (YITP-T-09-05) where he is benefitted from many stimulating discussions.
He is also grateful to the organizers of the workshop 
``The almost non-Gaussian universe" held at the Institut de Physique Theorique de Saclay and thank Leonard Senatore and Sebastien Renaux-Petel for useful discussions. SM is supported by JSPS. KK is supported by European Research Council, Research Councils UK and STFC.
\end{acknowledgments}

\appendix

\section{Optimal estimator}
In this section, we present the optimal estimator
to detect the trispectrum in equilateral type non-Gaussian models. As was shown in the main text, Eq.~(\ref{fiducial}) is a representative form of the trispectrum which has sufficiently large overlaps between trispectra in physically motivated models such as single field and multi-field DBI inflation.
Moreover, this trispectrum can be written in a factorisable form as
\begin{eqnarray}
T_{\zeta}(k_1,k_2,k_3,k_4) &=& \frac{g^{equil}_{\rm NL}}{k_1 k_2 k_3 k_4 (\frac{k_1+k_2+k_3+k_4}{4})^5} {\cal P}_\zeta^3, \\
&=& \frac{g^{equil}_{\rm NL} 4^5}{24} {\cal P}_\zeta^3  \int^{\infty}_{0} dt t^4
d(t,k_1) d(t,k_2) d(t,k_3) d(t,k_4),
\label{fiducialint}
\end{eqnarray}
where
\begin{equation}
d(k,t)=\frac{1}{k} \exp(-k t).
\end{equation}
Then the connected part of the trispectrum of the CMB temperature anisotropies is calculated as \begin{eqnarray}
\langle a_{l_1 m_1} a_{l_2 m_2} a_{l_3 m_3} a_{l_4 m_4} \rangle_c
&=&\frac{4^5}{24} \mathcal{P}_\zeta^3 \int d \Omega Y_{l_1 m_1}  Y_{l_2 m_2}  Y_{l_3 m_3}
 Y_{l_4 m_4}
\int t^4 dt \int r^2 dr
\gamma_{l_1}(r,t) \gamma_{l_2}(r,t) \gamma_{l_3}(r,t) \gamma_{l_4}(r,t)
\nonumber\\
&&\times  w_{l_1} w_{l_2} w_{l_3} w_{l_4},
\end{eqnarray}
where
\begin{equation}
\gamma_{l_i} (r, t)= \frac{2}{\pi} \int dk_i k_i^2  g_{T l_i}(k_i)
F(k_i,t) j_{l_i}(k_i, r),
\end{equation}
$g_{T l_i}$ is the radiative transfer function, $j_{l_i}$ is the spherical
Bessel function and $w_{l_i}$ is an experimental window function.
The optimal estimator is given by \cite{Regan:2010cn}
\begin{equation}
g_{\rm NL}^{equil} = \frac{S}{F},
\end{equation}
where
\begin{eqnarray}
S &=& \frac{1}{24} \sum_{l_i m_i}
\langle a_{l_1 m_1} a_{l_2 m_2} a_{l_3 m_3} a_{l_4 m_4} \rangle_c
\Big[
(C^{-1}a)_{l_1 m_1} (C^{-1}a)_{l_2 m_2} (C^{-1}a)_{l_3 m_3} (C^{-1}a)_{l_4 m_4}  \\
&& -6 (C^{-1})_{l_1 m_1 l_2 m_2} (C^{-1}a )_{l_3 m_3}(C^{-1}a)_{l_4 m_4}
+3 (C^{-1})_{l_1 m_1 l_2 m_2} (C^{-1})_{l_3 m_3 l_4 m_4},
\Big],
\end{eqnarray}
and
\begin{equation}
F= \frac{1}{24} \langle a_{l_1 m_1} a_{l_2 m_2} a_{l_3 m_3} a_{l_4 m_4} \rangle_c
\langle a_{l_1' m_1'} a_{l_2' m_2'} a_{l_3' m_3'} a_{l_4' m_4'} \rangle_c
(C^{-1})_{l_1 m_1, l_1' m_1'}(C^{-1})_{l_2 m_2, l_2' m_2'}(C^{-1})_{l_3 m_3, l_3' m_3'}
(C^{-1})_{l_4 m_4, l_4' m_4'}.
\end{equation}
Here
\begin{equation}
C_{l_i m_i, l_j m_j} =\langle a^*_{l_i m_i} a_{l_j m_j} \rangle,
\quad
(C^{-1}a )_{l_i m_i} = C^{-1}_{l_i m_i, l_j m_j}a_{l_j m_j}.
\end{equation}
Using the expression for the trispectrum (\ref{fiducialint}), the estimator for the equilateral
trispectrum can be written as
\begin{equation}
S = \frac{4^5 g_{\rm NL} ^{equil} {\cal P}_{\zeta}^3 }{24^2} \int t^4 dt \int r^2 dr \int d \Omega
\Big[
D(\Omega,r,t)^4 - 6 D(\Omega, r,t)^2 \langle D(\Omega,r,t)^2 \rangle_{\rm MC}
+3 \langle D(\Omega,r,t)^2 \rangle_{\rm MC}^2
\Big]
\end{equation}
where
\begin{eqnarray}
D(\Omega,r,t) &=& \sum_{l_i} w_{l_i} \gamma_{l_i}(r,t)
(C^{-1} a)_{{l_i} {m_i}} Y_{{l_i} {m_i}} \\
\langle D(\Omega,r,t)^2 \rangle_{\rm MC}
&=& \sum_{{l_i} {l_j}} w_{l_i} w_{l_j} \gamma_{l_i} \gamma_{l_j}
(C^{-1})_{{l_i} {m_i}, {l_j} {m_j}} Y_{{l_i} {m_i}} Y_{{l_j} {m_j}}.
\end{eqnarray}
The ensemble average can be evaluated using the Monte Carlo simulation.
The fisher error bound of $g_{\rm NL}^{equil}$ is given by $F^{-1}$ where
\begin{equation}
F = \sum_{L, l_i}
\frac{T^{l_1 l_2}_{l_3 l_4} (L)^2}{(2 L+1) C_{l_1} C_{l_2} C_{l_3} C_{l_4}},
\end{equation}
where we assume the covariant matrix is diagonal and the reduced trispectrum
is given by
\begin{equation}
T^{l_1 l_2}_{l_3 l_4} (L) = \frac{ 4^5 h_{l_1 L l_2} h_{l_3 L l_4} }{24}
\int t^4 dt \int r^2 dr
w_{l_1} w_{l_2} w_{l_3} w_{l_4}
\gamma_{l_1}(r,t) \gamma_{l_2}(r,t) \gamma_{l_3}(r,t) \gamma_{l_4}(r,t).
\end{equation}
Here $h_{l_i L l_j}$ is given by
\begin{equation}
h_{l_i L l_j}
= \sqrt{\frac{(2 l_i+1)(2 l_j+1)(2 L+1)}{4 \pi}}
\left(
\begin{array}{c c c}
l_i &L &l_j \\
0&0 &0
\end{array}
\right).
\end{equation}

\section{\label{parts}
Shape functions in general single field
k-inflation}

Here, based on our previous work \cite{Arroja:2009pd},
we summarise the shape functions for
the reduced trispectra in general single field
k-inflation described by the following action:
\begin{equation}
S=\frac{1}{2}\int d^4x\sqrt{-g}\left[R+2P(X,\phi)
\right],\label{action}
\end{equation}
where $\phi$ is the inflaton field,
$R$  is the Ricci scalar and
$X\equiv-(1/2)g^{\mu\nu}\partial_\mu\phi\partial_\nu\phi$, where
$g_{\mu\nu}$ is the metric tensor.

For this class of models, the third and the fourth order
interaction Hamiltonian of the field perturbation
$\delta \phi$ in the flat gauge at leading order
in the slow-roll expansion are given by
\begin{eqnarray}
H_I^{(3)}(\eta)&=&\int
 d^3x\left[Aa\delta\phi'^3+Ba\delta\phi'
\left(\partial\delta\phi\right)^2\right]\,,
\label{k_inf_third_Hamiltonian}\\
H_I^{(4)}(\eta)&=&\int d^3x\left[\beta_1\delta\phi'^4+
\beta_2\delta\phi'^2\left(\partial\delta\phi\right)^2
+\beta_3\left(\partial\delta\phi\right)^4\right]\,,
\label{k_inf_fourth_Hamiltonian}
\end{eqnarray}
where prime denotes derivative with respect to conformal
time $\eta$ and coefficients $A$, $B$, $\beta_1$, $\beta_2$
and $\beta_3$ are given by
\begin{equation}
A=-\frac{\sqrt{2X}}{2}\left(P_{,XX}+\frac{2}{3} X P_{,XXX}\right), \quad B=\frac{\sqrt{2X}}{2}P_{,XX}.
\end{equation}
\begin{eqnarray}
\beta_1&=&P_{,XX}\left(1-\frac{9}{8}c_s^2\right)-2X
 P_{,XXX}\left(1-\frac{3}{4}c_s^2\right)+\frac{X^3
 c_s^2}{P_{,X}}P_{,XXX}^2-\frac{1}{6} X^2 P_{,4X},
\nonumber\\
\beta_2&=&-\frac{1}{2}P_{,XX}\left(1-\frac{3}{2}c_s^2\right)+\frac{1}{2}X
 c_s^2P_{,XXX},
\nonumber\\
\beta_3&=&-\frac{c_s^2}{8}P_{,XX}.
\end{eqnarray}

Then, the shape function
$S^{k} _{\mathcal{T}}$ is composed of two parts
\begin{eqnarray}
S^{k} _{\mathcal{T}} =
S^{k (cont)} _{\mathcal{T}} +
S^{k (scalar)} _{\mathcal{T}}\,,
\end{eqnarray}
where $S^{k (cont)} _{\mathcal{T}}$
denotes the contribution from the contact interaction
and $S^{k (scalar)} _{\mathcal{T}}$
denotes that from the scalar exchange interaction,
respectively.

$S^{k (cont)} _{\mathcal{T}}$
is given by
\begin{eqnarray}
S^{k (cont)} _{\mathcal{T}} =
\left(-24 \beta_1 c_s ^3 S^{c_{1}} _{\mathcal{T}}
- \beta_2 c_s S^{c_{2}} _{\mathcal{T}}
- 2 \beta_3 c_s ^{-1} S^{c_{3}} _{\mathcal{T}} \right)
\frac{H^4}{4 X^2} N^8\,.
\label{shape_general_k_cont}
\end{eqnarray}
Here $S^{c_{1}} _{\mathcal{T}}$, $S^{c_{2}} _{\mathcal{T}}$
and $S^{c_{3}} _{\mathcal{T}}$ are the following
shape functions:
\begin{eqnarray}
S^{c_{1}} _{\mathcal{T}}
&=&
\frac{k_{12} \Pi_{i=1}^4k_i}{\left(\sum_{i=1}^4k_i\right)^5}
+ 3\;\; {\rm perms.}\,, \\
\label{shape_c1}
S^{c_{2}} _{\mathcal{T}}
 &=& \Biggl[\frac{k_{12}
 k_1^2k_2^2(\mathbf{k_3}\cdot\mathbf{k_4})}{\left(\sum_{i=1}^4k_i\right)^3\Pi_{i=1}^4k_i}\left(1+3\frac{(k_3+k_4)}{\sum_{i=1}^4k_i}+12\frac{k_3k_4}{\left(\sum_{i=1}^4k_i\right)^2}\right)
\nonumber\\
&&+\frac{k_{12} k_3^2k_4^2 (\mathbf{k_1}\cdot\mathbf{k_2})}
{\left(\sum_{i=1}^4k_i\right)^3\Pi_{i=1}^4k_i}\left(1+
3\frac{(k_1+k_2)}{\sum_{i=1}^4k_i}+
12\frac{k_1k_2}{\left(\sum_{i=1}^4k_i\right)^2}\right)
\Biggr]+3\;\; {\rm perms.}\,,
\label{shape_c2}\\
S^{c_{3}} _{\mathcal{T}}
&=& \frac{k_{12} (\mathbf{k_1}\cdot\mathbf{k_2})
(\mathbf{k_3}\cdot\mathbf{k_4})}
{\sum_{i=1}^4k_i\,\Pi_{i=1}^4k_i}
\left(1+\frac{\sum_{i<j}k_ik_j}{\left(\sum_{i=1}^4k_i\right)^2}+3\frac{\Pi_{i=1}^4k_i}{\left(\sum_{i=1}^4k_i\right)^3}\sum_{i=1}^4\frac{1}{k_i}+12\frac{\Pi_{i=1}^4k_i}{\left(\sum_{i=1}^4k_i\right)^4}\right)+3\;\; {\rm perms.}\,,
\label{shape_c3}
\end{eqnarray}
where ``$3\;\; {\rm perms.}$" denotes the permutations
$(k_1 \leftrightarrow k_2)$,
$(k_3 \leftrightarrow k_4)$ and
$(k_1 \leftrightarrow k_2, k_3 \leftrightarrow k_4)$.
In Eq.~(\ref{shape_general_k_cont}), $N=H/\sqrt{2 P_{,X} c_s}$.

Similarly, $S^{k (scalar)} _{\mathcal{T}}$
is given by
\begin{eqnarray}
S^{k (scalar)} _{\mathcal{T}} =
\left(A^2 c_s^4  S^{s_{1}} _{\mathcal{T}}
+ AB c_s^2 S^{s_{3}} _{\mathcal{T}}
+B^2 S^{s_{2}} _{\mathcal{T}} \right)
\frac{c_s^2 H^2 N^{10}}{8X^2}\,.
\label{shape_general_k_scalar}
\end{eqnarray}
Here $S^{s_{1}} _{\mathcal{T}}$, $S^{s_{2}} _{\mathcal{T}}$
and $S^{s_{3}} _{\mathcal{T}}$ are the following
shape functions:
\begin{eqnarray}
S^{s_1} _{\mathcal{T}}
 &=& -9 k_{12}
(k_1 k_2 k_3 k_4)^{1/2}
\biggl[ \tilde{\mathcal{F}}_1
(k_1,k_2,-k_{12},k_3,k_4,k_{12}) -
\tilde{\mathcal{F}}_1(-k_1,-k_2,-k_{12},k_3,k_4,k_{12})
\nonumber\\&&
+ \tilde{\mathcal{F}}_1
(k_3,k_4,-k_{12},k_1,k_2,k_{12}) -
\tilde{\mathcal{F}}_1(-k_3,-k_4,-k_{12},k_3,k_4,k_{12})
\biggr]+3\;\; {\rm perms.}\,,
\label{shape_s1}
\end{eqnarray}
\begin{eqnarray}
S^{s_{2}} _{\mathcal{T}} &=&
S^{s_{2a}} _{\mathcal{T}} +
S^{s_{2b}} _{\mathcal{T}} +
S^{s_{2c}} _{\mathcal{T}} +
S^{s_{2d}} _{\mathcal{T}}\,,
\label{def_shape_s2}\\
S^{s_{2a}} _{\mathcal{T}}
&=& -k_{12} (k_1 k_2 k_3 k_4)^{1/2}
(\mathbf{k_1}\cdot\mathbf{k_2})
(\mathbf{k_3}\cdot\mathbf{k_4})
\biggl[ \tilde{\mathcal{F}}_2
(-k_{12},k_1,k_2,k_{12},k_3,k_4) -
\tilde{\mathcal{F}}_2(-k_{12},-k_1,-k_2,k_{12},k_3,k_4)
\nonumber\\
&&+
 \tilde{\mathcal{F}}_2
(-k_{12},k_3,k_4,k_{12},k_1,k_2) -
\tilde{\mathcal{F}}_2(-k_{12},-k_3,-k_4,k_{12},k_1,k_2)
\biggr]+3\;\; {\rm perms.}\,,
\label{shape_s2a}
\\
S^{s_{2b}} _{\mathcal{T}}
 &=& -2 k_{12} (k_1 k_2 k_3 k_4)^{1/2}
 (\mathbf{k_1}\cdot\mathbf{k_2})
(\mathbf{k_{12}}\cdot\mathbf{k_4})
\biggl[ \tilde{\mathcal{F}}_2
(-k_{12},k_1,k_2,k_3,k_4,k_{12}) -
\tilde{\mathcal{F}}_2(-k_{12},-k_1,-k_2,k_3,k_4,k_{12})
\nonumber\\
&&+\tilde{\mathcal{F}}_2
(k_3,k_4,-k_{12},k_{12},k_1,k_2) -
\tilde{\mathcal{F}}_2(-k_3,-k_4,-k_{12},k_{12},k_1,k_2)
\biggr]+3\;\; {\rm perms.}\,,
\label{shape_s2b}
\\
S^{s_{2c}} _{\mathcal{T}}
 &=& 2 k_{12} (k_1 k_2 k_3 k_4)^{1/2}
 (\mathbf{k_{12}}\cdot\mathbf{k_2})
(\mathbf{k_3}\cdot\mathbf{k_4})
\biggl[ \tilde{\mathcal{F}}_2
(k_1,k_2,-k_{12},k_{12},k_3,k_4) -
\tilde{\mathcal{F}}_2(-k_1,-k_2,-k_{12},k_{12},k_3,k_4)
\nonumber\\
&&+\tilde{\mathcal{F}}_2
(-k_{12},k_3,k_4,k_1,k_2,k_{12}) -
\tilde{\mathcal{F}}_2(-k_{12},-k_3,-k_4,k_1,k_2,k_{12})
\biggr]+3\;\; {\rm perms.}\,,
\label{shape_s2c} \\
S^{s_{2d}} _{\mathcal{T}}
 &=& 4 k_{12} (k_1 k_2 k_3 k_4)^{1/2}
(\mathbf{k_{12}}\cdot\mathbf{k_2})
(\mathbf{k_{12}}\cdot\mathbf{k_4})
\biggl[ \tilde{\mathcal{F}}_2
(k_1,k_2,-k_{12},k_3,k_4,k_{12}) -
\tilde{\mathcal{F}}_2(-k_1,-k_2,-k_{12},k_3,k_4,k_{12})
\nonumber\\
&&+
\tilde{\mathcal{F}}_2
(k_3,k_4,-k_{12},k_1,k_2,k_{12}) -
\tilde{\mathcal{F}}_2(-k_3,-k_4,-k_{12},k_1,k_2,k_{12})
\biggr]+3\;\; {\rm perms.}\,,
\label{shape_s2d}
\end{eqnarray}
\begin{eqnarray}
S^{s_{3}} _{\mathcal{T}} &=&
S^{s_{3a}} _{\mathcal{T}} +
S^{s_{3b}} _{\mathcal{T}} +
S^{s_{3c}} _{\mathcal{T}} +
S^{s_{3d}} _{\mathcal{T}}\,,
\label{def_shape_s3}\\
S^{s_{3a}} _{\mathcal{T}}
 &=& 3 k_{12} (k_1 k_2 k_3 k_4)^{1/2}
 (\mathbf{k_3}\cdot\mathbf{k_4})
\biggl[ \tilde{\mathcal{F}}_3
(k_1,k_2,-k_{12},k_{12},k_3,k_4) -
\tilde{\mathcal{F}}_3(-k_1,-k_2,-k_{12},k_{12},k_3,k_4)
\nonumber\\
&&+\tilde{\mathcal{F}}_4
(-k_{12},k_3,k_4,k_1,k_2,k_{12}) -
\tilde{\mathcal{F}}_4(-k_{12},-k_3,-k_4,k_1,k_2,k_{12})
\biggr]+3\;\; {\rm perms.}\,,
\label{shape_s3a} \\
S^{s_{3b}} _{\mathcal{T}}
 &=& 6 k_{12} (k_1 k_2 k_3 k_4)^{1/2}
 (\mathbf{k_{12}}\cdot\mathbf{k_4})
\biggl[ \tilde{\mathcal{F}}_3
(k_1,k_2,-k_{12},k_3,k_4,k_{12}) -
\tilde{\mathcal{F}}_3(-k_1,-k_2,-k_{12},k_3,k_4,k_{12})
\nonumber\\
&&+\tilde{\mathcal{F}}_4
(k_3,k_4,-k_{12},k_1,k_2,k_{12}) -
\tilde{\mathcal{F}}_4(-k_3,-k_4,-k_{12},k_1,k_2,k_{12})
\biggr]+3\;\; {\rm perms.}\,,
\label{shape_s3b}\\
S^{s_{3c}} _{\mathcal{T}}
 &=& 3 k_{12} (k_1 k_2 k_3 k_4)^{1/2}
 (\mathbf{k_1}\cdot\mathbf{k_2})
\biggl[ \tilde{\mathcal{F}}_4
(-k_{12},k_1,k_2,k_3,k_4,k_{12}) -
\tilde{\mathcal{F}}_4(-k_{12},-k_1,-k_2,k_3,k_4,k_{12})
\nonumber\\
&&+\tilde{\mathcal{F}}_3
(k_3,k_4,-k_{12},k_{12},k_1,k_2) -
\tilde{\mathcal{F}}_3(-k_3,-k_4,-k_{12},k_{12},k_1,k_2)
\biggr]+3\;\; {\rm perms.}\,,
\label{shape_s3c}\\
S^{s_{3d}} _{\mathcal{T}}
 &=& -6 k_{12} (k_1 k_2 k_3 k_4)^{1/2}
 (\mathbf{k_{12}}\cdot\mathbf{k_2})
\biggl[ \tilde{\mathcal{F}}_4
(k_1,k_2,k_{-12},k_3,k_4,k_{12}) -
\tilde{\mathcal{F}}_4(-k_1,-k_2,k_{-12},k_3,k_4,k_{12})
\nonumber\\
&&+\tilde{\mathcal{F}}_3
(k_3,k_4,-k_{12},k_1,k_2,k_{12}) -
\tilde{\mathcal{F}}_3(-k_3,-k_4,-k_{12},k_1,k_2,k_{12})
\biggr]+3\;\; {\rm perms.} \,,
\label{shape_s3d}
\end{eqnarray}
where again ``$3\;\; {\rm perms.}$" denotes the permutations
$(k_1 \leftrightarrow k_2)$,
$(k_3 \leftrightarrow k_4)$ and
$(k_1 \leftrightarrow k_2, k_3 \leftrightarrow k_4)$.
Here we have defined four $\tilde{\mathcal{F}}_i$
 functions (with $i=1,\ldots,4$) as follows;
\begin{eqnarray}
\tilde{\mathcal{F}}_1(k_1,k_2,k_3,k_4,k_5,k_6)
&=&-4|k_1 k_2 k_3 k_4 k_5 k_6|^\frac{1}{2}
\frac{1}{\mathcal{A}^3\mathcal{C}^3}\left(1+3\frac{\mathcal{A}}{\mathcal{C}}+6\frac{\mathcal{A}^2}{\mathcal{C}^2}\right)\,,\\
\tilde{\mathcal{F}}_2(k_1,k_2,k_3,k_4,k_5,k_6)
&=&-\frac{|k_1k_4|^\frac{1}{2}}{|k_2k_3k_5k_6|^\frac{3}{2}}\frac{1}{\mathcal{A}\mathcal{C}}
\bigg[
      1+\frac{k_5+k_6}{\mathcal{A}}+2\frac{k_5k_6}{\mathcal{A}^2}
      \nonumber\\&&
      +\frac{1}{\mathcal{C}}  \left(k_2+k_3+k_5+k_6+\frac{1}{\mathcal{A}}\left(\left(k_2+k_3\right)\left(k_5+k_6\right)+2k_5k_6\right)+2\frac{k_5k_6\left(k_2+k_3\right)}{\mathcal{A}^2}\right)
      \nonumber\\&&
      +\frac{2}{\mathcal{C}^2}\bigl(k_5k_6+\left(k_2+k_3\right)\left(k_5+k_6\right)+k_2k_3+\frac{1}{\mathcal{A}}\left(k_2k_3\left(k_5+k_6\right)+2k_5k_6\left(k_2+k_3\right)\right)\nonumber\\&&+2\frac{k_2k_3k_5k_6}{\mathcal{A}^2}\bigr)
 +\frac{6}{\mathcal{C}^3}\left(k_2k_3\left(k_5+k_6\right)+k_5k_6\left(k_2+k_3\right)+2\frac{k_2k_3k_5k_6}{\mathcal{A}}\right)
\nonumber\\
&&+24\frac{k_2k_3k_5k_6}{\mathcal{C}^4}
\bigg]\,,\\
\tilde{\mathcal{F}}_3(k_1,k_2,k_3,k_4,k_5,k_6)
&=&2\frac{|k_1k_2k_3k_4|^\frac{1}{2}}{|k_5k_6|^\frac{3}{2}}\frac{1}{\mathcal{A}\mathcal{C}^3}
\left[
      1+\frac{k_5+k_6}{\mathcal{A}}+2\frac{k_5k_6}{\mathcal{A}^2}
      +\frac{3}{\mathcal{C}}\left(k_5+k_6+2\frac{k_5k_6}{\mathcal{A}}\right)
      +12\frac{k_5k_6}{\mathcal{C}^2}
\right]\,,\\
\tilde{\mathcal{F}}_4(k_1,k_2,k_3,k_4,k_5,k_6)
&=&2\frac{|k_1k_4k_5k_6|^\frac{1}{2}}{|k_2k_3|^\frac{3}{2}}\frac{1}{\mathcal{A}^3\mathcal{C}}
\bigg[1+\frac{\mathcal{A}}{\mathcal{C}}+\frac{\mathcal{A}^2}{\mathcal{C}^2}+\frac{k_2+k_3}{\mathcal{C}}+2\frac{\mathcal{A}\left(k_2+k_3\right)+k_2k_3}{\mathcal{C}^2}
\nonumber\\&&
\qquad\qquad\qquad\qquad\qquad\quad
+3\frac{\mathcal{A}}{\mathcal{C}^3}\left(\mathcal{A}\left(k_2+k_3\right)+2k_2k_3\right)+12k_2k_3\frac{\mathcal{A}^2}{\mathcal{C}^4}\bigg]\,,
\end{eqnarray}
where $\mathcal{A}$ is defined by the sum of the last three arguments of
the $\tilde{\mathcal{F}}_i$ functions as  $\mathcal{A}=k_4+k_5+k_6$ and
$\mathcal{C}$ is defined by the sum of all the arguments as
$\mathcal{C}=k_1+k_2+k_3+k_4+k_5+k_6$.

In Tables~\ref{table_shape_correlation_trispectrum},
we summarise the correlations among these shape functions for the full configurations,
the configurations with $\epsilon=1$, the equilateral
configurations ($\epsilon=1$, $\alpha=\gamma=0$),
respectively.
\begin{widetext}
\begin{center}
\begin{table} [h!]
\begin{tabular} { |c|c|c|c|c|c|c|c|c| }
\hline
 & $c_1$ & $c_2$ & $c_3$
&$s_1$  & $s_2$
&$s_3$ & $\tilde{s}_2$& $\tilde{s}_3$   \\
\hline
Overlap-full& $1.00$ & $-0.55$  & $0.35$
& $0.98$ &$0.83$
&  $-0.89$ &$-0.056$&$0.24$ \\
\hline
$\epsilon=1$& $1.00$ &$-0.64$  & $0.72$
&$1.00$  & $0.86$
& $-0.92$ &$0.32$&$0.17$  \\
\hline
$equilateral$& $1.00$ &$-0.72$ & $0.79$
& $0.98$ & $0.90$
& $-0.93$ &$0.48$&$0.44$  \\
\hline
\end{tabular}
\caption{Shape correlations between the equilateral shape
and the shapes of primordial trispectra
in general single field k inflation. The correlations between
the shapes that appear in multi-field inflation models are also
shown in the last two columns.}
\label{table_shape_correlation_trispectrum}
\end{table}
\end{center}
\end{widetext}

It is worth noting that the following properties
\begin{eqnarray}
F(S^{c_1} _{\mathcal{T}},
\; S_{\mathcal{T}}) &=& \sum_i a_i
F(S^{c_1} _{\mathcal{T}},\;
S^{i} _{\mathcal{T}})\,,\\
F(S_{\mathcal{T}},
\; S_{\mathcal{T}})
&=& \sum_{i,j} a_i a_j
F(S^{i} _{\mathcal{T}},\;
S^{j} _{\mathcal{T}})\,,
\label{selfcorr_decomp}
\end{eqnarray}
hold for the shape function given by
\begin{eqnarray}
S_{\mathcal{T}}  = \sum_i a_i
S^{i} _{\mathcal{T}}\,,
\label{corr_decomp}
\end{eqnarray}
where $i=c_1, c_2, c_3, s_1, s_2, s_3$ and $a_i$'s
are corresponding coefficients.
By combining the shape correlations
obtained in Table~\ref{table_shape_correlation_trispectrum}
and the properties shown above,
we can calculate correlations
of the shape functions in any general
single field k-inflation models
against the equilateral shape (\ref{shape_c1}),
once the action (\ref{action}) is specified.

Especially, in the case of single field DBI inflation,
the coefficients in the Hamiltonians
(\ref{k_inf_third_Hamiltonian}) and
(\ref{k_inf_fourth_Hamiltonian}) are given by
\begin{eqnarray}
&&A=-\frac{1}{2 \dot{\phi} c_s^5}\,,\;\;\;\;
B=\frac{1}{2 \dot{\phi} c_s^3}\,\nonumber\\
&&\beta_1 = \frac{1}{2 c_s^7 \dot{\phi}^2}\,,\;\;\;\;
\beta_2=\frac{1}{4 c_s^3 \dot{\phi}^2}\,,\;\;\;\;
\beta_3 = -\frac{1}{8 c_s \dot{\phi}^2}\,.
\end{eqnarray}

In multi-field DBI inflation model \cite{Mizuno:2009mv},
in addition to the shape functions
$S^{c_{1}} _{\mathcal{T}}$, $S^{c_{2}} _{\mathcal{T}}$,
$S^{c_{3}} _{\mathcal{T}}$, $S^{s_{1}} _{\mathcal{T}}$
$S^{s_{2}} _{\mathcal{T}}$, $S^{s_{3}} _{\mathcal{T}}$,
we find it convenient to define
the following shape functions
$S^{\tilde{s}_{2}} _{\mathcal{T}}$ and
$S^{\tilde{s}_{3}} _{\mathcal{T}}$ given by
\begin{eqnarray}
S^{\tilde{s}_{2}} _{\mathcal{T}} &=&
S^{s_{2a}} _{\mathcal{T}} -
S^{s_{2b}} _{\mathcal{T}} -
S^{s_{2c}} _{\mathcal{T}} +
S^{s_{2d}} _{\mathcal{T}}\,,
\label{def_shape_s2_tilde}\\
S^{\tilde{s}_{3}} _{\mathcal{T}} &=&
S^{s_{3a}} _{\mathcal{T}} -
S^{s_{3b}} _{\mathcal{T}} +
S^{s_{3c}} _{\mathcal{T}} -
S^{s_{3d}} _{\mathcal{T}}\,.
\label{def_shape_s3_tilde}
\end{eqnarray}
The table \ref{table_shape_correlation_trispectrum} shows the shape correlations
between these shapes and the equilateral shape. It is clear that these shapes have
significantly low correlations which explain the reason why the final correlation
between the shapes in the equilateral model and the multi-field DBI models is lower
than the single field model.

\section{Bispectrum estimation for single field DBI inflation
\label{sec_bispectrum_estimation}}

In this section, we explain our method to compare the amplitudes of the
bispectrum using the overlap integration.
We introduce the shape function $S_B$
\begin{eqnarray}
S_B (k_1, k_2, k_3) = k_1 ^2 k_2 ^2 k_3^2
B_\zeta (k_1, k_2, k_3)\,,
\label{def_shapefunc_bispectrum}
\end{eqnarray}
for the bispectrum of the curvature
perturbation $ B_\zeta (k_1, k_2, k_3)$
defined by
\begin{eqnarray}
\langle \zeta({\bf k_1}) \zeta({\bf k_2}) \zeta({\bf k_3})
\rangle = (2 \pi)^3 \delta^{(3)}
({\bf k_1} + {\bf k_2} + {\bf k_3})  B_\zeta (k_1, k_2, k_3)\,.
\label{def_pri_bispectrum}
\end{eqnarray}

In \cite{Chen:2006nt},
it was shown that at leading order in the slow roll expansion
and small sound speed limit, shape function for the bispectrum
in the single field DBI inflation is given by
\begin{eqnarray}
S^{DBI} _{B} = {\cal P}_\zeta^2 \frac{4}{c_s^2 \Pi_i k_i}
\left(-\frac{1}{K} \sum_{i > j} k_i^2 k_j^2 + \frac{1}{2K^2}
\sum_{i \neq j} k_i^2 k_j^3 + \frac{1}{8} \sum_i k_i^3\right)
\,,\label{shape_func_b_dbi_theory}
\end{eqnarray}
where $K=k_1 + k_2 + k_3$.

The shape (\ref{shape_func_b_dbi_theory})
is not factorisable and it is not easy to perform
an optimal analysis using CMB observations. Thus
a factorisable shape which approximates
(\ref{shape_func_b_dbi_theory}) was proposed by
\cite{Babich:2004gb} which is given by
\begin{eqnarray}
S^{equil} _B = {\cal P}_\zeta^2 \frac{18}{5} f_{\rm NL} ^{equil}
\Pi_i k_i^2 \left(-\frac{1}{k_1^3 k_2^3}
- \frac{1}{k_1^3 k_3^3}-  \frac{1}{k_2^3 k_3^3}
-\frac{2}{k_1^2 k_2^2 k_3^2} + \frac{1}{k_1 k_2^2 k_3^3}
+5\;\; {\rm perms.}\right)\,,
\label{shape_func_b_equil}
\end{eqnarray}
where the permutations act only on the last term
in parentheses.

One way to relate $f_{\rm NL}^{equil}$ to the prediction of DBI inflation
is to compare the amplitude of the shape function for the equilateral
configuration $k_1=k_2=k_3$. We get
\begin{eqnarray}
f_{\rm NL} ^{equil} = -\frac{35}{108 c_s^2}
\simeq - 0.32 c_s^{-2}\,.
\label{fnl_dbi_conv}
\end{eqnarray}
In fact, in \cite{Fergusson:2008ra}, these two shapes are shown to be $99\%$ correlated
to each other based on the following primordial 
shape correlator for two different shape functions
$S_B$ and $S_{B'}$
\begin{eqnarray}
\bar{\mathcal{C}}_B (S_B, S_{B'}) = \frac{F_B (S_B, S_{B'})}
{\sqrt{F_B (S_B, S_B) F_B (S_{B'}, S_{B'})}}\,,
\label{shape_correlator_b}
\end{eqnarray}
which is constructed from
\begin{eqnarray}
F_B(S_B, S_{B'}) = \int d \mathcal{U}_k S_B(k_1, k_2, k_3)
S_{B'} (k_1, k_2, k_3) w_B(k_1, k_2, k_3)\,.
\label{pribispectrum_correlator_part}
\end{eqnarray}
In Eq.~(\ref{pribispectrum_correlator_part})
the integration is performed for the region
where the triangle condition for $(k_1, k_2, k_3)$ holds
and weight function $w_B$ is given by
\begin{eqnarray}
w_B= \frac{1}{k_1 + k_2 + k_3}\,.
\end{eqnarray}

For these two slightly different shapes, it would be
enough to match the amplitudes of the bispectra evaluated at the
equilateral configuration ($k_1=k_2=k_3$) where the amplitude
have a peak by setting the relation (\ref{fnl_dbi_conv}).
However, it is more appropriate to match the amplitude
and shape of the bispectra by taking into account all possible configurations.
In this context, we use a different way to estimate the amplitude
of the bispectrum in single field DBI inflation
using the information of all possible configurations.
For this purpose, we rewrite
Eq.~(\ref{shape_func_b_dbi_theory}) in the following form:
\begin{eqnarray}
S^{DBI} _{B} = -11 {\cal P}_{\zeta}^2
\frac{f_{\rm NL} ^{equil}}
{\bar{\mathcal{C}}_B (S^{DBI} _{B}, S^{equil} _B)}
\frac{1}{\Pi_i k_i}
\left(-\frac{1}{K} \sum_{i > j} k_i^2 k_j^2 + \frac{1}{2K^2}
\sum_{i \neq j} k_i^2 k_j^3 + \frac{1}{8} \sum_i k_i^3\right)
\,,\label{shape_func_b_dbi_est}
\end{eqnarray}
where the numerical factor
in Eq.~(\ref{shape_func_b_dbi_est}) is chosen so that when
$f_{\rm NL} ^{equil}=1$ and
$\bar{\mathcal{C}}_B (S^{DBI} _{B}, S^{equil} _B) =1$,
the following conditions are satisfied,
\begin{eqnarray}
F_B(S^{DBI} _{B}, S^{DBI} _{B})
&=& F_B(S^{equil} _{B},S^{equil} _{B})\,,
\nonumber\\
F_B(S^{DBI} _{B},S^{equil} _{B}) &>&0\,.
\label{def_sdbibuni}
\end{eqnarray}

Of course, $\bar{\mathcal{C}}_B (S^{DBI} _{B}, S^{equil} _B) =1$
is not true in reality and this factor will
enhance the amplitude of the signal
for a given $f_{\rm NL} ^{equil}$.
Then, by comparing Eqs.~(\ref{shape_func_b_dbi_theory})
with (\ref{shape_func_b_dbi_est}), we can relate
 $f_{\rm NL} ^{equil}$ with $c_s^2$ as
\begin{eqnarray}
f_{\rm NL} ^{equil} = -0.36 c_s^{-2}\,,
\label{fnl_dbi_ours}
\end{eqnarray}
where we have used ${\cal P}_\zeta = H^4/(2 \dot{\phi}^2)$
for single field DBI inflation.
As is expected, this gives almost the same value as
the one given by Eq.~(\ref{fnl_dbi_ours}),
due to the fact that there is a large overlap
between the two shapes. However, for the trispectrum the difference between
the two approaches tends to be large as the trispectrum has five parameters
even assuming the scale invariance and thus matching the amplitude at a specific
point in the five-dimensional parameter space is generally not enough to ensure that
we get the same signal. In this case, it is more appropriate to use all the shape
information using the overlap integration to compare the shape and amplitude of
trispectra.


\end{document}